\documentclass[a4paper,12pt]{article}
\usepackage{latexsym}
\usepackage{amssymb,amsmath,amsthm}
\font\gross=cmr12 scaled \magstep3
\font\mittel=cmr12 scaled \magstep1

\font\Mittel=cmbx12 scaled \magstep1

\newcommand{\beq}{\begin{eqnarray*}}
\newcommand{\beqn}{\begin{eqnarray}}
\newcommand{\eeq}{\end{eqnarray*}}
\newcommand{\eeqn}{\end{eqnarray}}
\newcommand{\bitem}{\begin{enumerate}}
\newcommand{\eitem}{\end{enumerate}}
\newcommand{\bmatr}{\begin{array}}
\newcommand{\ematr}{\end{array}}

\newcommand{\up}{\uparrow}
\newcommand{\down}{\downarrow}
\newcommand{\ts}{\textstyle}
\newcommand{\ds}{\displaystyle}
\newcommand{\la}{\langle}
\newcommand{\ra}{\rangle}
\newcommand{\pt}{\partial}
\newcommand{\pro}{\mathop\Pi}

\newcommand{\bk}{{\bf k}}

\newcommand{\p}{{\bf p}}
\newcommand{\q}{{\bf q}}

\newcommand{\M}{{\cal M}}
\newcommand{\N}{{\cal N}}

\newcommand{\vep}{\varepsilon}
\newcommand{\vp}{\varphi}
\newcommand{\vph}{\varphi}

\begin{document}
\include{epsf}
{
\thispagestyle{empty}
\baselineskip=16pt
$$ $$
\vskip 3cm
\centerline{\gross Resummation of Feynman Diagrams } 
\smallskip
\centerline{\gross and the Inversion of Matrices  }
\bigskip
\bigskip
\centerline{by}
\bigskip
\bigskip
\centerline{{\mittel Detlef Lehmann}\footnote{ e-mail: 
 lehmann@math.tu-berlin.de}}
\centerline{\mittel Technische Universit\"at Berlin}
\centerline{\mittel Fachbereich Mathematik Ma 7-2} 
\centerline{\mittel Sta{\ss}e des 17. Juni 136}
\centerline{\mittel D-10623 Berlin, GERMANY}
\vskip 4.5cm 
\noindent{\bf Abstract:} In many field theoretical models one has to
resum two- and four-legged subdia\-grams in order to determine their 
behaviour. In this article we present a novel formalism which does this 
in a nice way. It is based on the central limit theorem of probability 
 and an inversion formula for matrices which is obtained 
 by repeated application of the Feshbach projection method. We discuss 
applications to the Anderson model, to the many-electron system and 
 to the $\vp^4$-model. In particular, for the many-electron system 
 with attractive delta-interaction,   
we find that the existence of a BCS gap and a macroscopic value 
of the Hubbard-Stratonovich field for zero momentum enforce 
each other.

\bigskip
\bigskip
\vfill
\eject  }
%
%
%
%
\pagenumbering{arabic}
\baselineskip=16pt

\section{Introduction}
\bigskip
\indent
The computation of correlation functions in field theoretical models is a difficult 
 problem. In this article we present a novel approach which applies to models 
where a two point function can be written as 
$$ S(x,y)=\int [P+Q]_{x,y}^{-1}\> d\mu(Q)\>. \eqno (1.1) $$
Here $P$ is some operator diagonal in momentum space, typically determined by the 
unperturbed Hamiltonian, and $Q$ is diagonal in coordinate space. 
The functional integral is taken with respect to some probability measure 
 $d\mu(Q)$ and goes over the matrix elements of $Q$. $[\;\cdot\;]_{x,y}^{-1}$ 
denotes the $x,y$-entry of the matrix $[P+Q]^{-1}$. Our starting point 
is always a model in finite volume and with positive lattice spacing in 
which case the operator $P+Q$ and the functional integral in (1.1) 
  becomes 
huge- but finite-dimensional. In the end we take the infinite volume limit 
 and, if wanted, the continuum limit. 
\par
Our treatment is based on the following identity which is obtained by
repeated application of the Feshbach formula (Lemma 3.2 below). 
 It  is proven in Theorem 3.3. 
 Let $B=(B_{kp})_{k,p\in\M}\in \mathbb C^{N\times N}$, $\M$ some 
 index set, $|\M|=N$ and let 
$$G(k):=\left[ B^{-1}\right]_{kk}  \eqno (1.2)$$
Then one has
$$G(k)={1\over\ds B_{kk}+\sum_{r=2}^N (-1)^{r+1}
   \!\!\!\!\!\!\!\!\!
    \sum_{p_2\cdots p_{r}\in\M\setminus\{k\}\atop p_i\ne p_j}
  \!\!\!\!\!\!\!  
  B_{kp_2}G_k(p_2)B_{p_2 p_3}\cdots B_{p_{r-1} p_r} 
  G_{kp_2\cdots p_{r-1}}(p_r)B_{p_r k}  }\eqno (1.3)$$
where 
 $ G_{k_1\cdots k_j}(p)=\left[ (B_{st})_{s,t\in \M\setminus 
    \{k_1\cdots k_j\}}\right]^{-1}_{pp}$ 
is the $p,p$ entry of the inverse of the matrix which is obtained
from $B$ by deleting the rows and columns labelled by  
  $k_1,\cdots,k_j$. In Section 2  we apply this formula to a matrix of 
 the form $B=$ self adjoint + $i\vep\,{I\!d}$, which, 
  for $\vep\ne 0$, has 
 the property that all submatrices $ (B_{st})_{s,t\in \M\setminus 
    \{k_1\cdots k_j\}}$ are invertible. 
\par
There is also a formula for the off-diagonal inverse matrix 
 elements. It reads 
$$\left[B^{-1}\right]_{kp}=-G(k) B_{kp} G_k(p)+
  \sum_{r=3}^N(-1)^{r+1} \!\!\!\!\!\!\!  
 \!\!\! \sum_{t_3\cdots t_r\in\M\setminus\{k,p\}\atop t_i\ne t_j}
  \!\!\!\!\!\! G(k) B_{kt_3} G_k(t_3) B_{t_3t_4} 
  \cdots B_{t_rp}G_{kt_3\cdots t_r}(p)  \eqno (1.4)$$
These formulae also hold in the case where the matrix $B$ has a 
block structure $B_{kp}=(B_{k\sigma,p\tau})$ where, say, $\sigma,\tau\in 
 \{\up,\down\}$ are some spin variables. 
   In that case the $B_{kp}$ are small matrices, 
the $G_{k_1\cdots k_j}(p)$ are matrices of the same size 
 and the 1/$\cdot$ in (1.3) 
means inversion of matrices, see Theorem 3.3 below. 
\par
 The paper is organized as follows. 
In the next section we demonstrate the method by applying  it 
  to the averaged Green function of the
Anderson model. The Schwinger-Dyson equation for that model reads 
 $G^{-1}=G_0^{-1}+\Sigma(G_0)$  where  $\Sigma(G_0)$ is the sum of 
 all  two-legged one-particle irreducible diagrams. 
Application of (1.3) 
 leads to an integral equation $G^{-1}=G_0^{-1}+
 \sigma(G)$ where 
$\sigma(G)$ is the sum of all two-legged graphs without two-legged 
 subgraphs. The latter equation has two advantages. First, 
 $\Sigma$ is the sum of one-particle irreducible diagrams, but 
 these diagrams may very well have two-legged subdiagrams and 
usually these are the diagrams which produce anomalously large 
 contributions. 
And second, the propagator for $\sigma(G)$ is the interacting two 
 point function $G$, which, for the Anderson model, is more regular 
than 
the free two point function $G_0$ which is the propagator for the 
 diagrams contributing to $\Sigma(G_0)$. More precisely, 
 the series for  
 $\sigma(G)$ can be expected to be asymptotic, that is, its 
 lowest order contributions are a good approximation if the 
 coupling is small, but, usually, the series for $\Sigma(G_0)$ 
 is not asymptotic.
\par
For the many-electron system and for the $\vp^4$ model 
repeated application of (1.3,4) amounts to a resummation of two- and 
four-legged subgraphs. This is discussed in section 4. In  
section 5 we discuss how our method is related to the 
integral equations which can be found in the literature.
  The proof  of the inversion formula is given in section 3. 
\bigskip
\bigskip
%
%
%
%
\section{Application to the Anderson Model}
\bigskip
\indent
Let coordinate space be a lattice of finite volume with periodic 
 boundary conditions, lattice spacing $1/M$ and volume 
 $[0,L]^d$: 
$$\ts \Gamma=\left\{ x={1\over M}(n_1,\cdots,n_d)\;|\;
 0\le n_i\le ML-1\right\}=
 \left( {1\over M}\Bbb Z\right)^d/(L\Bbb Z)^d  \eqno (2.1)$$
Momentum space is given by 
$$\ts \M:=\Gamma^\sharp=\left\{k={2\pi\over L}(m_1,\cdots,m_d)\;|\; 
    0\le m_i\le ML-1\right\}
 =\left( {2\pi\over L}\Bbb Z\right)^d/(2\pi M \Bbb Z)^d \eqno (2.2)$$
We consider the averaged Green function of the Anderson model given
by 
$$\la G\ra(x,x'):=\int \left[ -\Delta-z+\lambda V
     \right]_{x,x'}^{-1} 
  dP(V) \eqno (2.3)$$
where the random potential is Gaussian,  
$$dP(V)= \pro_{x\in\Gamma} \ts e^{-{V_x^2\over 2}} {dV_x\over
       \sqrt{2\pi}}.  \eqno(2.4)$$
Here $z=E+i\vep$ and $\Delta$ is the discrete Laplacian, 
$$\left[ -\Delta-z+\lambda V
     \right]_{x,x'}= -M^2\sum_{i=1}^d 
   \left( \delta_{x',x+ e_i/M} 
  +\delta_{x',x- e_i/M}-2\delta_{x',x}\right) -z \,\delta_{x,x'}
  +\lambda V_x \,
  \delta_{x,x'} \eqno (2.5)$$
By taking the Fourier transform, one has 
\setcounter{equation}{5}
\label{sec2}
\beqn
 \la G\ra(x,x')&=&{\ts {1\over M^d L^d}} \sum_{k\in\M} 
  e^{ik(x'-x)} \la G\ra (k) \label{2.6} \\
 \la G\ra (k)&=& \int_{\Bbb R^{N^d}} \ts
   \left[ a_k\delta_{k,p}+{\lambda\over \sqrt{N^d}} v_{k-p}
  \right]^{-1}_{k,k}\;dP(v) \label{2.7}
\eeqn
where $N^d=(ML)^d=|\Gamma|=|\M|$, $dP(v)$ is given by (2.10) or (2.11) 
 below, depending on whether $N^d$ is even or odd, and 
$$a_k=4M^2\sum_{i=1}^d \sin^2\left[\ts {k_i\over 2M}\right] 
  -E-i\vep \eqno (2.8)$$
 The rigorous control of $\la G\ra (k)$ for small 
 disorder $\lambda$ and energies inside the spectrum of the unperturbed 
 Laplacian, $E\in [0,4M^2]$, in  which case $a_k$ has a root if $\vep\to 0$, 
 is still an open problem [AG,K,MPR,P,W]. It is expected 
 that $\lim_{\vep\searrow 0}\lim_{L\to\infty} 
 \la G\ra (k)=1/(a_k-\sigma_k)$ where Im$\sigma=O(\lambda^2)$. 

The integration variables $v_{q}$ in (2.7) are given by the discrete 
Fourier transform of the $V_x$. In particular, 
observe that, if $F$ denotes the unitary matrix of discrete 
 Fourier transform, the variables  
$$v_{q} \equiv(FV)_{q}={\ts {1\over \sqrt{N^d}}}\sum_{x\in\Gamma} 
   e^{-iqx} V_x={\ts \left({M\over L}\right)^{d\over2}  \;
  {1\over M^d}} \sum_{x\in\Gamma} 
   e^{-iqx} V_x \equiv  {\ts \left({M\over L}\right)^{d\over2}}
  \, \hat V_q  \eqno(2.9)$$
would not have a limit if $V_x$ would be deterministic and cutoffs are 
removed, since the $\hat V_q$ are the quantities which have a limit
in that case. But since the $V_x$ are integration variables, we 
choose a unitary transform to keep the integration measure
invariant. Observe also that $v_q$ is complex, 
 $v_q=u_q+iw_q$. Since $V_x$ is real, $u_{-q}=u_q$ and 
 $w_{-q}=-w_q$. In order to transform $dP(V)$ to momentum space, 
we have to choose a set $\M^+\subset\M$ 
  such that either $q\in \M^+$ or $-q\in \M^+$. If $N$ is odd, 
  the only momentum with  
 $q=-q$ or $w_q=0$ is $q=0$. In that case $dP(V)$ becomes 
$$ dP(v)=e^{-{u_{0}^2\over2}}
 {\ts 
  {du_{0}\over \sqrt{2\pi}}} \pro_{q\in \M^+} 
  e^{-(u_q^2+w_q^2)} 
 {\ts {du_q dw_q\over \pi}} \eqno (2.10) $$
For even $N$ we get 
$$ dP(v) =e^{-{1\over2}(u_{0}^2+u_{q_0}^2)}
 {\ts 
  {du_{0}du_{q_0}\over {2\pi}}} \pro_{q\in \M^+} 
  e^{-(u_q^2+w_q^2)} 
 {\ts {du_q dw_q\over \pi}}   \eqno (2.11) $$
where $q_0={2\pi m\over L}$ is the unique nonzero momentum  for 
which  ${2\pi\over L}m=2\pi M 
 (1,\cdots,1)- {2\pi\over L}m$. 

\vspace{0.6cm}
Now we apply the inversion formula (1.3) to the inverse matrix 
 element in 
$$\la G\ra (k)= \int_{\Bbb R^{N^d}} \ts
   \left[ a_k\delta_{k,p}+{\lambda\over \sqrt{N^{d}}} v_{k-p}
  \right]^{-1}_{k,k}\;dP(v) \eqno (2.7) $$
We start with the `two loop approximation',  
 which we define by retaining only the $r=2$ term in the 
 denominator of the right hand side of (1.3),
$$G(k)\approx 
  {1\over B_{kk}-\sum_{p\in\M\setminus\{k\}} 
    B_{kp} G_k(p) B_{pk} }  \eqno(2.12)$$
Thus, let 
$$\ts G(k):=\left[ a_k\delta_{k,p}+{\lambda\over \sqrt{N^{d}}} v_{k-p}
  \right]^{-1}_{k,k}=G(k;v,z) \eqno (2.13)$$
In the infinite volume limit the spacing $2\pi/L$ of the momenta becomes 
arbitrary small. Hence, in computing an inverse matrix element, 
it should not matter whether a single column and row labelled by some 
  momentum $t$ is absent or not. 
 In other words, in the infinite volume limit one should have 
$$ \phantom{ {\;\;\;{\rm for}\;\;L\to\infty} }
    G_t(p)=G(p)\;\;\;{\rm for}\;\;L\to\infty  \eqno (2.14) $$
and similarly $G_{t_1\cdots t_j}(p)=G(p)$ as long as $j$ is independent of the 
volume. We remark however that if the matrix has a block structure, say 
 $B=(B_{k\sigma,p\tau})$ with $\sigma,\tau\in\{\up,\down\}$ some spin variables,
 this structure has to be respected. That is, for a given momentum $k$ 
 all rows and columns labelled by $k\!\up,\;k\!\down$ 
   have to be eliminated, since otherwise (2.14) may not be 
 true. 
\par
Thus the two loop approximation gives 
$$ G(k)={1\over a_k+{\lambda\over \sqrt{N^d}} v_0 
  -{\lambda^2\over N^d}\sum_{p\ne k} v_{k-p}\,G(p)\, v_{p-k} }
   \eqno (2.15)$$
For large $L$, we can disregard the ${\lambda\over \sqrt{N^d}} v_0$ 
 term. Introducing $\sigma_k=\sigma_k(v,z)$ according to 
$$G(k)=:{1\over a_k-\sigma_k}\;,  \eqno (2.16)$$
we get 
$$\sigma_k={\ts {\lambda^2\over N^d}}\sum_{p\ne k} {|v_{k-p}|^2 
  \over a_p-\sigma_p }
 \approx {\ts {\lambda^2\over N^d}}\sum_{p} {|v_{k-p}|^2 
  \over a_p-\sigma_p } \eqno (2.17)$$
and arrive at 
$$\la G\ra (k)=\int {1\over \ds a_k-{\ts {\lambda^2\over N^d}} 
  \sum_{p}^{\phantom{I}} {|v_{k-p}|^2\over a_p- 
  {\ts {\lambda^2\over N^d}} 
  \sum_{t} {|v_{p-t}|^2\over a_{t}-{\lambda^2\over N^d}\Sigma\cdots}  }  }
  \;dP(v)  \eqno (2.18)$$
Now consider the infinite volume limit $L\to\infty$ or $N=ML\to\infty$. 
   By the central limit theorem of probability 
 ${1\over \sqrt{N^d}}\sum_q\left(|v_q|^2-\la |v_q|^2\ra\right)$ is, 
as a sum of independent 
 random variables, normal distributed. Note that only 
 a prefactor of $1/\sqrt{N^d}$ is required for that property. 
In particular, 
if $F$ is some bounded function independent of $N$, 
 sums which come with a prefactor 
 of $1/N^d$ like 
  ${1\over N^d}\sum_q c_q |v_q|^2$ can be substituted by
their expectation value,
$$\lim_{N\to\infty}\int F\Bigl({\ts{1\over N^d}}
  \sum_kc_k |v_k|^2\Bigr)dP(v)=F\Bigl(\lim_{N\to\infty}{\ts{1\over N^d}}
  \sum_kc_k \la |v_k|^2\ra\Bigr) \eqno (2.19)$$
Therefore, in the two loop approximation, one obtains in the infinite 
 volume limit
$$\la G\ra (k)= {1\over \ds a_k-{\ts {\lambda^2\over N^d}} 
  \sum_{p}^{\phantom{I}} {\la |v_{k-p}|^2\ra \over a_p- 
  {\ts {\lambda^2\over N^d}} 
  \sum_{t} {\la |v_{p-t}|^2\ra \over a_{t}
  -{\lambda^2\over N^d}\Sigma \cdots}  }  }
  \;=:\; {1\over a_k-\la \sigma_k \ra }  \eqno (2.20)$$ 
where the quantity $\la \sigma_k \ra$ satisfies the integral 
 equation 
$$\la \sigma_k \ra= {\ts {\lambda^2\over N^d}}\sum_{p}{
   {\la |v_{k-p}|^2 \ra \over a_p-\la \sigma_p\ra }\;  }
  \buildrel L\to\infty\over \to \;
   {\ts {\lambda^2\over M^d}}\int_{[0,2\pi M]^d}  
  { {d^dp\over (2\pi)^d}} \,
   {\la |v_{k-p}|^2\ra  \over a_p-\la\sigma_p\ra }  \eqno (2.21)$$
For a Gaussian distribution $\la |v_q|^2\ra=1$ for all $q$ 
such that $\la \sigma_k\ra =\la\sigma\ra$ becomes 
 independent of $k$. Thus we end up with 
$$\la G\ra(k)={1\over 4M^2\sum_{i=1}^d\sin^2\left[ {k_i\over 2M}
  \right] -E-i\vep-\la\sigma\ra} \eqno (2.22)$$
where $\la\sigma\ra$ is a solution of
\setcounter{equation}{22}
\beqn
  \la\sigma\ra&=&
    {\ts {\lambda^2\over M^d}}\int_{[0,2\pi M]^d}
   { {d^dp\over (2\pi)^d}} 
\, {1\over 4M^2\sum_{i=1}^d
  \sin^2\left[{p_i\over 2M}\right]-z-\la\sigma\ra}\nonumber \\
 &=&\lambda^2\int_{[0,2\pi]^d}
   { {d^dp\over (2\pi)^d}} 
\, {1\over 4M^2\sum_{i=1}^d
  \sin^2\left[{p_i\over 2}\right]-z-\la\sigma\ra}\>.  
\eeqn
This equation is of course well known and one deduces from it that
it generates a small imaginary part Im$\,\sigma=O(\lambda^2)$ if 
the energy $E$ is within the spectrum of $-\Delta$. 
       
\bigskip
We now add the higher loop terms (the terms for $r>2$ in the  
 denominator of (1.3)) to our discussion and give an 
interpretation in terms of Feynman diagrams. To make the volume
factors more explicit, asume that the lattice spacing in 
 coordinate space is $1/M=1$ such that $N=L$. 
\par
For the Anderson model, Feynman graphs may be obtained by brutally 
expanding 
\beqn
  \lefteqn{\int 
   \bigl[a_k\delta_{k,p}+{\ts{\lambda\over \sqrt L^{d}} v_{k-p}}
  \bigr]^{-1}_{k,k}\;dP =\sum_{r=0}^\infty 
   \int \left( C[VC]^r\right)_{kk} \,dP}  \\
 && =\sum_{r=0}^\infty  {\ts {(-\lambda)^r\over \sqrt{L^d}^r}}    
  \sum_{p_2\cdots p_{r}} {\ts {1\over a_ka_{p_2}\cdots a_{p_{r}} 
  a_k}} \int v_{k-p_2} v_{p_2-p_3}\cdots v_{p_{r}-k} \,dP 
   \nonumber
\eeqn
For a given $r$, this may be represented as 
 in figure 1 ($c_k:=1/a_k$).

\begin{figure}[thb]
 \centerline{\epsfbox{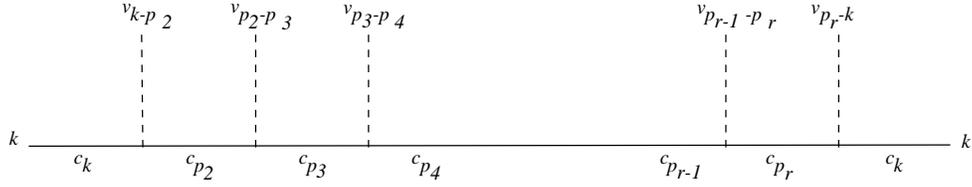}}
 \caption{A string of particle lines with unpaired squiggles (dashed lines)}
 \label{Figure 1}
\end{figure} 

\noindent
The integral over the $v$ gives a sum of $(r-1)!!$ terms where 
 each term is a product of $r/2$ Kroenecker-delta's, the terms for odd $r$ 
 vanish.  
If this is substituted in (2.24), the number of independent momentum
sums is cut down to $r/2$ and  each of the $(r-1)!!$ terms may be
represented by a diagram

\begin{figure}[thb]
 \centerline{\epsfbox{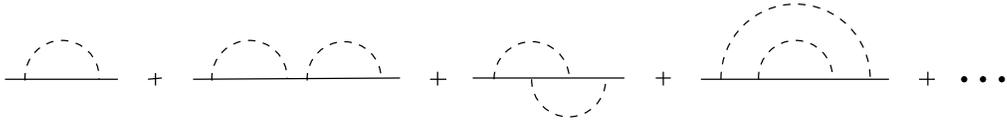}}
 \caption{Lowest order diagrams contributing to (2.24) 
   for $r=2$ and $r=4$ }
 \label{Figure 2}
\end{figure}
\noindent
where, say, the value of the third diagram is given by 
 ${\lambda^4\over L^{2d}}\sum_{p_1,p_2} {1\over a_k
 a_{k+p_1} a_{k+p_1+p_2} a_{k+p_2}a_k}$. 
 For short: 
$$\la G\ra (k)={\rm sum\;of\; all\; two\; legged\; diagrams\,.}
  \eqno (2.25)$$
Since the value of a diagram depends on its subgraph structure, 
one distinguishes, in the easiest case, two types of diagrams. 
Diagrams with or without two-legged subdiagrams.  Those diagrams 
with two-legged subgraphs usually produce anomalously large 
 contributions. They are devided further into the one-particle  
 irreducible ones and the reducible ones. Thereby a diagram is 
 called one-particle reducible if it can be made disconnected 
 by cutting one solid or `particle' line (no squiggle or dashed line), see 
 also figure 3.  

\begin{figure}[thb]
 \centerline{\epsfbox{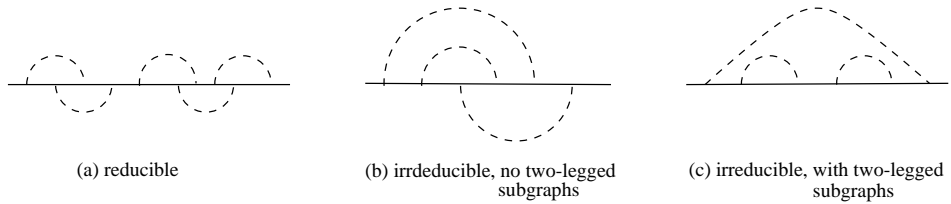}}
 \caption{Reducible and irreducible diagrams}
 \label{Figure 3}
\end{figure} 
\noindent
The reason for introducing reducible and irreducible diagrams is 
 that the reducible ones can be easily resummed by writing down 
the Schwinger-Dyson equation which states that if 
the self energy $\Sigma_k$ is defined through 
$$\la G\ra (k)={1\over a_k-\Sigma_k(G_0)} \eqno (2.26)$$
then $\Sigma_k(G_0)$ is the sum of all 
 amputated (no $1/a_k$'s at the ends) one particle irreducible 
diagrams. Here we wrote $\Sigma_k(G_0)$ to indicate that the factors
(`propagators') assigned to the solid lines of the diagrams
contributing to $\Sigma_k$ are given by the free two point function 
 $G_0(p)={1\over a_p}$. However, the diagrams 
contributing to $\Sigma_k(G_0)$  still  
contain anomalously large contributions, namely irreducible 
graphs which contain two-legged subgraphs like diagram (c) in  
 figure 3. 

In the following we show, using the inversion formula (1.3) 
 including all higher loop terms, that all graphs with two-legged 
subgraphs can be eliminated or resummed by writing down 
the following integral equation for $\la G\ra$:
$$ \la G\ra (k)= {1\over a_k-\sigma_k(\la G\ra)} \eqno (2.27) $$
where 
\begin{quote}
 $\sigma_k(\la G\ra)$ 
  is the sum of all amputated two-legged diagrams 
 which do not contain any two-legged subdiagrams, but now 
 with propagators $\la G\ra(k)={1\over a_k-\sigma_k}$ 
instead of  $G_0={1\over a_k}$
\end{quote}
\noindent
which may be formalized as in (2.35) below. 
 The advantage of this formula is that the series for  
 $\sigma_k(\la G\ra)$ can be expected to be asymptotic, that is, its 
 lowest order contributions are a good approximation if the 
 coupling is small, but, usually, the series for $\Sigma_k(G_0)$ 
 is not asymptotic. Thus, 
in order to rigorously controll $\la G\ra (k)$,   one has to 
define a suitable space of propagators, to estimate the sum of all 
 amputated two-legged graphs without two-legged subgraphs on 
 that space and then finally to show that the equation (2.27) has a 
 solution on this space. We intend to address this problem in another
paper. 

\bigskip
We now show (2.27) for the Anderson model. For fixed $v$ one has 
$$G(k,v)={1\over a_k-\sigma_k(v)}  \eqno (2.28)$$
where 
$$\sigma_k(v)=\sum_{r=2}^{L^d}(-1)^{r}\!\!\!\!\!\!
    \sum_{p_2\cdots p_{r}
 \atop p_i\ne p_j,\> p_i\ne k}
  \!\! \ts {\lambda^r\over \sqrt{ L^d}^r} 
 \, G_k(p_2)\cdots G_{kp_2\cdots p_{r-1}}(p_r)\, v_{k-p_2}
  \cdots v_{p_r-k}  \eqno (2.29)$$
We cutoff the $r$-sum in (2.29) at some arbitrary but fixed order $\ell<L^d$ 
where $\ell$ is choosen to be independent of the volume. 
 Furthermore we substitute $G_{kp_2\cdots p_j}(p)$ by $G(p)$. 
Thus   
$$\la G\ra (k)=\biggl\la {1\over a_k-\sum_{r=2}^\ell 
   \sigma_k^r(v) } \biggr\ra  \eqno (2.30) $$
where 
$$\sigma_k^r(v)=(-1)^r{\ts {\lambda^r\over \sqrt{ L^d}^r}} \!\!\!\!
  \sum_{p_2\cdots p_{r} \atop p_i\ne p_j,\> p_i\ne k} \!\!\!\!
  G(p_2)\cdots G(p_r)\, v_{k-p_2} \cdots v_{p_r-k}\eqno (2.31)$$
Consider first two strings $s_k^{r_1}$, $s_k^{r_2}$ where 
$$ s_k^r(v)={\ts {\lambda^r\over \sqrt{ L^d}^r}} \!\!\!\!
  \sum_{p_2\cdots p_{r} \atop p_i\ne p_j,\> p_i\ne k} \!\!\!\!
   c^r_{kp_2\cdots p_r}\, v_{k-p_2} \cdots v_{p_r-k}\eqno (2.32)$$
and the $c^r_{kp_2\cdots p_r}$ are some numbers. Then in the
infinite volume limit 
$$\la s_k^{r_1} s_k^{r_2} \ra= \la s_k^{r_1}\ra\, 
  \la s_k^{r_2} \ra  \eqno (2.33)$$ 
because all pairings which connect the two strings have an extra 
 volume factor $1/L^d$. Namely, if the two strings are 
  disconnected, there
are $(r_1+r_2)/2$ loops and a volume factor of 
   $1/\sqrt{L^d}^{(r_1+r_2)}$ giving $(r_1+r_2)/2$ 
 Riemannian sums. If the two strings are connected, there are only 
 $(r_1+r_2-2)/2$ loops leaving an extra factor of $1/L^d$. By the 
 same argument one has in the infinite volume limit
$$\la (s_k^{r_1})^{n_1}\cdots (s_k^{r_m})^{n_m}\ra = 
  \la s_k^{r_1} \ra^{n_1}\cdots \la s_k^{r_m}\ra^{n_m}\eqno (2.34)$$
which results in 
$$\la G\ra(k)= {1\over \ds a_k -\sum_{r=2}^\ell 
 {\ts {(-\lambda)^r\over \sqrt{ L^d}^{\,r}}}\!\!\!\!
  \sum_{p_2\cdots p_{r} \atop p_i\ne p_j,\> p_i\ne k} \!\!\!\!
  \la G\ra(p_2)\cdots \la G\ra (p_r)\, \la v_{k-p_2} \cdots 
   v_{p_r-k} \ra} \eqno (2.35) $$
The condition $p_2,\cdots,p_r\ne k$ and $p_i\ne p_j$ means exactly 
 that two-legged subgraphs are forbidden. Namely, for a 
two-legged subdiagram  
as in (c) in figure 3, the incomming and outgoing momenta  
$p$, $p'$ (to which are
assigned propagators $\la G\ra(p)$, $\la G\ra(p')$)
must be equal which is forbidden by 
 the condidtion $p_i\ne p_j$ in (2.35).

However, we cannot take the limit $\ell\to\infty$ in (2.35) since the 
 series in the denominator of (2.35) is only an asymptotic one. 
 To see this a bit more clearly suppose for the moment that there 
 were no restrictions on the momentum sums. Then, if
$V=({\lambda\over\sqrt{L^d}}v_{k-p})_{k,p}$ and 
  $\la G\ra=(\la G\ra (k)\,\delta_{k,p})_{k,p}$, 
$${\ts {\lambda^r\over \sqrt{ L^d}^r}} 
   \sum_{p_2\cdots p_r}\la G\ra(p_2)\cdots \la G\ra(p_r)\,
  \la v_{k-p_2} \cdots v_{p_r-k} \ra 
 =\left\la (V[\la G\ra V]^{r-1})_{kk} \right\ra \eqno (2.36)$$ 
and for $\ell \to \infty$ we would get 
$$\la G\ra (k)={1\over  a_k-\la (V {\la G\ra V\over Id +\la G\ra V} 
    )_{kk} \ra }
 = {1\over  a_k-\la (V {1\over \la G\ra^{-1} + V} V
    )_{kk} \ra }   \eqno (2.37)$$
That is, the factorials produced by the number of diagrams in the
denominator of (2.35) are basically the same as those in the 
 expansion 
$$ \int_{\Bbb R}{\ts {x^2\over z+\lambda x}}\, e^{-{x^2\over 2}} 
  {\ts {dx\over\sqrt{2\pi}}}\> =\>\sum_{r=0}^\ell \ts 
  {\lambda^{2r}\over z^{2r+1}}\,(2r+1)!!\>
 +R_{\ell+1}(\lambda)\, \eqno (2.38)$$ 
where the remainder satisfies the bound  
$|R_{\ell+1}(\lambda)|\le \ell!\,const_{\! z}^\ell\, \lambda^\ell $. 

\bigskip
We close this section with two  further remarks. 
So far the computations were done in momentum space. One may wonder 
 what one gets if the inversion formula (1.3) is applied to 
 $[-\Delta+z+\lambda V]^{-1}$ in coordinate space. Whereas a 
 geometric series expansion of $[-\Delta+z+\lambda V]^{-1}$ gives 
 a representation in terms of the simple random walk, 
  application of (1.3) results in a representation in terms 
 of the self avoiding walk: 
$$[-\Delta+z+\lambda V]_{0,x}^{-1}=\sum_{\gamma:0\to x\atop 
  \gamma\;{\rm self\;avoiding}}
 {\det\left[ (-\Delta+z+\lambda V)_{y,y'\in\Gamma\setminus\gamma}
   \right] \over\det\left[ (-\Delta+z+\lambda V)_{y,y'\in\Gamma}
      \right]  }  \eqno (2.39)$$
where $\Gamma$ is the lattice in coordinate space.  Namely, if $|x|>1$, the 
inversion formula (1.4) for the off-diagonal elements gives 
\beq
 \lefteqn{[-\Delta+\lambda V]_{0,x}^{-1}=\sum_{r=3}^{L^d}(-1)^{r+1} 
 \!\!\!\!\!\!\!\!\! 
  \sum_{x_3\cdots x_r\in\Gamma\setminus\{0,x\}\atop x_i\ne x_j}
  \!\!\!\!\!\!\!\! G(0)G_0(x_3)\cdots G_{0x_3\cdots x_r}(x)
  \,(-\Delta)_{0 x_3}
  \cdots (-\Delta)_{x_r x}  }  \\
 &&=\sum_{r=3}^{L^d}  \sum_{x_2=0,x_3,\cdots ,x_r,x_{r+1}=x\in\Gamma\atop
    |x_i-x_{i+1}|=1\; \forall i=2\cdots r}
  \!\!\!\! {\det\left[ (-\Delta+\lambda V)_{y,y'\in\Gamma\setminus\{0\}}
   \right] \over\det\left[ (-\Delta+\lambda V)_{y,y'\in\Gamma}
      \right]  } \cdots 
   {\det\left[ (-\Delta+\lambda V)_{y,y'\in\Gamma\setminus\{0,x_3\cdots x_r,x\}}
   \right] \over\det\left[
   (-\Delta+\lambda V)_{y,y'\in\Gamma\setminus\{0,x_3\cdots x_r\}}
      \right]  }  
\eeq
which coincides with (2.39). 

Finally we remark that, while the argument following (2.32) leads to a 
factorization
property for on-diagonal elements in momentum space, 
 $\la G(k)\,G(p)\ra=\la G(k)\ra\,\la G(p)\ra$, 
 there is no such property for products of off-diagonal elements 
which appear in a quantity like 
$$\Lambda(q)= {\ts {1\over L^d}}\sum_{k,p}\ts \left\la 
 \left[ a_k\delta_{k,p}+{\lambda\over \sqrt L^{\,d}} v_{k-p}
  \right]^{-1}_{k,p}
  \left[ \bar a_k\delta_{k,p}+{\lambda\over \sqrt L^{\,d}} \bar v_{k-p}
  \right]^{-1}_{k-q,p-q} \right\ra  \eqno (2.40) $$ 
which is the Fourier transform of 
$\bigl\la \left| [-\Delta+z+\lambda V]^{-1}_{x,y}
   \right|^2\bigr\ra$. (Each off-diagonal inverse matrix element is proportional 
 to $1/\sqrt{L^d}$, therefore the prefactor of $1/L^d$ in (2.40) is correct.)
\bigskip
\bigskip
%
%
%
%
\section{Proof of the Inversion Formula}
\label{sec3}

\bigskip
\noindent{\bf Lemma 3.1:} Let $B\in\Bbb C^{k\times n}$, 
 $C\in\Bbb C^{n\times k}$ and let $Id_k$ denote the identity 
 in $\Bbb C^{k\times k}$. 
Then:
\bitem
\item[{\bf (i)}] $Id_k-BC\;\;{\rm invertible}\;\;\;
  \Leftrightarrow\;\;\;Id_n-CB\;\;{\rm invertible}$. 
\item[{\bf (ii)}]  If the left or the right hand side of (i) fullfilled, 
 then
 $\ts C\,{1\over Id_k-BC} =\ts  {1\over Id_n-CB}\,C $.
\eitem

\bigskip
\noindent{\bf Proof:}  Let 
$$B=\left(\bmatr{ccc}
  - & \vec b_1 & - \\  
      & \vdots & \\ 
    - & \vec b_k & - 
\ematr\right),\;\;\;\;
  C=\left(\bmatr{ccc}
   |& &|\\
   \vec c_1&\cdots &\vec c_k\\
   |& &|
\ematr\right)  $$ 
where the $\vec b_j$ are $n$-component row vectors and the 
 $\vec c_j$ are $n$-component column vectors. Let $\vec x\in 
 {\rm Kern}(Id-CB)$.  Then $\vec x=CB \vec x=\sum_j \lambda_j \vec c_j$ if 
 we define $\lambda_j:= (\vec b_j,\vec x)$. Let 
 $\vec \lambda=(\lambda_j)_{1\le j\le k}$. Then $[(Id-BC)\vec\lambda]_i
  =\lambda_i-\sum_j(\vec b_i,\vec c_j)\lambda_j 
  =(\vec b_i,\vec x)-\sum_j(\vec b_i,\vec c_j)\lambda_j=0$ 
 since $\vec x=\sum_j \lambda_j \vec c_j$, thus 
$\vec \lambda \in {\rm Kern}(Id-BC)$. On the other hand, if 
 some $\vec\lambda\in {\rm Kern}(Id-BC)$, then $\vec x:=\sum_j\lambda_j 
 \vec c_j\in {\rm Kern}(Id-CB)$ which proves (i). Part (ii) then follows 
from 
$ C\ts={1\over Id_n-CB}\,(Id_n-CB)C 
  ={1\over Id_n-CB}\,C (Id_k-BC)$ $\blacksquare$

\bigskip
\bigskip
The inversion formula (1.3,4) is obtained by iterative application of 
 the next lemma, which states the Feshbach formula for finite dimensional 
matrices. For a more general version one may look in [BFS], Theorem 2.1. 

\bigskip
\bigskip
\noindent{\bf Lemma 3.2:} Let 
 $\ts h=\left({A\atop C}\;{B\atop D}\right)\in\Bbb C^{n\times n}$ where 
$A\in \Bbb C^{k\times k}$, 
 $D\in \Bbb C^{(n-k)\times (n-k)}$ are invertible and  
 $B\in \Bbb C^{k\times (n-k)}$, $C\in\Bbb C^{(n-k)\times k}$. Then 
$$ h\;\;{\rm invertible}\;\;\;\;
  \Leftrightarrow\;\;\;\;A-BD^{-1}C\;\;{\rm invertible}\;\;\;\;
  \Leftrightarrow\;\;\;\; D-CA^{-1}B\;\;{\rm invertible} \eqno (3.1)$$
and if one of the conditions in (3.1) is fullfilled, one has 
 $h^{-1} = \left({E\atop G}\;{F\atop H}\right)$ 
where 
\setcounter{equation}{1}
\beqn
\ts E={1\over A-BD^{-1}C}\>,&\hskip 0.7cm&\ts
      H={1\over D-CA^{-1}B}\>, \\
  F=-EBD^{-1}=-A^{-1}BH\>,&\hskip 0.7cm& G=-HCA^{-1}=-D^{-1}CE\>. 
\eeqn

\bigskip
\noindent{\bf Proof:} We have, using Lemma 3.1 in the second line, 
\beq
  A-BD^{-1}C\;\;{\rm inv.}&\Leftrightarrow &
    Id_k-A^{-1}BD^{-1}C\;\;{\rm inv.}  \\
 & \Leftrightarrow &
    Id_{n-k}-D^{-1}CA^{-1}B\;\;{\rm inv.}  \\
 & \Leftrightarrow & D-CA^{-1}B\;\;{\rm inv.}  
\eeq
Furthermore, again by Lemma 3.1, 
$$\ts D^{-1}C\,{1\over Id-A^{-1}BD^{-1}C}=
    {1\over Id-D^{-1}CA^{-1}B}\,D^{-1}C=
  {1\over D-CA^{-1}B}\,C=HC$$
and 
$$\ts A^{-1}B\,{1\over Id-D^{-1}CA^{-1}B}=
  {1\over Id-A^{-1}BD^{-1}C}\,A^{-1}B=
 {1\over A-BD^{-1}C}\,B=EB$$
which  proves the last equalities in (3.3), 
 $HCA^{-1}=D^{-1}CE$
and 
 $EBD^{-1}= A^{-1}BH $. 
Using these equations and the definition of $E,F,G$ and $H$ one computes 
$$\ts \left({A\atop C}\;{B\atop D}\right)
      \left({E\atop G}\;{F\atop H}\right)
     = \left({E\atop G}\;{F\atop H}\right)
      \left({A\atop C}\;{B\atop D}\right)
  =\left({Id\atop 0}\;{0\atop Id}\right)$$
It remains to show that the invertibility of $h$ implies the invertibility 
 of $A-BD^{-1}C$. To this end let $P=
  \left({Id\atop \phantom{C}}\;{\phantom{B}\atop 0}\right)$, 
 $\bar P=
  \left({0\atop \phantom{C}}\;{\phantom{B}\atop Id}\right)$ such that 
$A-BD^{-1}C=PhP-Ph\bar P(\bar P h\bar P)^{-1} \bar P h P$. Then 
\beq
  (A-BD^{-1}C)Ph^{-1}P&=&PhPh^{-1}P-
   Ph\bar P(\bar P h\bar P)^{-1} \bar P h Ph^{-1}P \\
 &=&Ph(1-\bar P)h^{-1}P-
   Ph\bar P(\bar P h\bar P)^{-1} \bar P h(1- \bar P)h^{-1}P \\
 &=&P-Ph\bar P h^{-1} P +Ph \bar P h^{-1} P=P  
\eeq
and similarly $Ph^{-1}P(A-BD^{-1}C)=P$ which proves the invertibility 
  of $A-BD^{-1}C$ $\blacksquare$

\bigskip
\bigskip
\noindent{\bf Theorem 3.3:} Let $B\in \Bbb C^{nN\times nN}$ be 
 given by  $B=(B_{kp})_{k,p\in\M}$, $\M$ some index set, $|\M|=N$, 
 and  $B_{kp}=
 (B_{k\sigma,p\tau})_{\sigma,\tau\in I}\in \Bbb C^{n\times n}$ 
  where $I$ is another index set, $|I|=n$. 
 Suppose that $B$ and, for any $\N\subset \M$, the 
  submatrix $(B_{kp})_{k,p\in\N}$ is invertible. 
For $k\in \M$ let 
$$G(k):=\left[B^{-1}\right]_{kk}\in \Bbb C^{n\times n}\eqno (3.4)$$ 
and, if $\N\subset\M$, $k\notin \N$, 
$$G_{\N}(k):= \left[\{(B_{st})_{s,t\in\M\setminus\N}\}^{-1}\right]_{kk}
    \in \Bbb C^{n\times n} \eqno (3.5)$$ 
Then one has 
\bitem
\item[\bf (i)] The on-diagonal block matrices of $B^{-1}$ are given by 
$$G(k)={1\over\ds B_{kk}-\sum_{r=2}^N (-1)^{r}
   \!\!\!\!\!\!\!\!\!\!\!
    \sum_{p_2\cdots p_{r}\in\M\setminus\{k\}\atop p_i\ne p_j}
  \!\!\!\!\!\!\! \!\! 
  B_{kp_2}G_k(p_2)B_{p_2 p_3}\cdots B_{p_{r-1} p_r} 
  G_{kp_2\cdots p_{r-1}}(p_r)B_{p_r k}  }  \eqno (3.6)$$
where $1/\;\cdot\;$ is inversion of $n\times n$ matrices. 
\item[\bf (ii)] Let $k,p\in\M$, $k\ne p$. Then the 
  off-diagonal block matrices of $B^{-1}$ can be expressed in terms 
 of the $G_{\N}(s)$ and the $B_{st}$, 
$$[B^{-1}]_{kp}=-G(k) B_{kp} G_k(p)-
  \sum_{r=3}^N(-1)^{r} \!\!\!\!\!\!\!  
 \!\!\! \sum_{t_3\cdots t_r\in\M\setminus\{k,p\}\atop t_i\ne t_j}
  \!\!\!\!\!\! G(k) B_{kt_3} G_k(t_3) B_{t_3t_4} 
  \cdots B_{t_rp}G_{kt_3\cdots t_r}(p) \eqno (3.7) $$
\eitem

\bigskip
\noindent{\bf Proof:} Let $k$ be fixed and let $p,p'\in\M\setminus\{k\}$ 
 below label columns and rows.  By Lemma 3.2 we have 
$$ \left(\bmatr{cccc}
  G(k) & && \\ & && \\ &&*& \\ &&& 
\ematr\right)\;=\;\left( 
 \bmatr{cccc} B_{kk} & \;-&B_{kp}&-\; \\
  | & && \cr B_{p'k}&&B_{p'p}& \\ |&&& 
 \ematr\right)^{-1}
 =\;\;
\left(\bmatr{cccc}
  E &\;- &F&-\; \\ |& && \\ G&&H& \\ |&&& 
\ematr\right)  $$
where 
\setcounter{equation}{7}
\beqn
  G(k)=E&
   =&{1\over\ds B_{kk}-\sum_{p,p'\ne k}B_{kp}\left[\{(B_{p' p})_{ 
  p',p\in\M\setminus\{k\}}\}^{-1}\right]_{pp'} B_{p'k} }  \\
 &=&{1\over \ds B_{kk}-\sum_{p\ne k}B_{kp} G_k(p) B_{pk}-
  \sum_{p,p'\ne k\atop p\ne p'} B_{kp}\left[\{ (B_{p' p})_{ 
  p',p\in\M\setminus\{k\}}\}^{-1}\right]_{pp'} B_{p'k} }\;\; \nonumber
\eeqn 
and 
\beqn F_{kp}=\left[ B^{-1}\right]_{kp}&=& 
  -G(k)\sum_{t\ne k} B_{kt} 
 \left[\{ (B_{p' p})_{p',p\in\M\setminus\{k\}}\}^{-1}\right]_{tp} \\
 &=&-G(k)B_{kp}G_k(p)-G(k)\sum_{t\ne k,p} B_{kt} 
 \left[\{ (B_{p' p})_{p',p\in\M\setminus\{k\}}\}^{-1}\right]_{tp}\nonumber
\eeqn
Apply Lemma 3.2 now to the matrix 
 $\{ (B_{p' p})_{p',p\in\M\setminus\{k\}}\}^{-1}$ and proceed by 
 induction to obtain after $\ell$ steps 
\beqn
  G(k)&=&{1\over\ds B_{kk}-\sum_{r=2}^\ell (-1)^{r}
   \!\!\!\!\!\!\!\!\!\!\!
    \sum_{p_2\cdots p_{r}\in\M\setminus\{k\}\atop p_i\ne p_j}
  \!\!\!\!\!\!\! \!\! 
  B_{kp_2}G_k(p_2)B_{p_2 p_3}\cdots B_{p_{r-1} p_r} 
  G_{kp_2\cdots p_{r-1}}(p_r)B_{p_r k} -R_{\ell+1}  } \\
 F_{kp}&=&-G(k) B_{kp} G_k(p)-
  \sum_{r=3}^\ell (-1)^{r} \!\!\!\!\!\!\!  
 \!\!\! \sum_{t_3\cdots t_r\in\M\setminus\{k,p\}\atop t_i\ne t_j}
  \!\!\!\!\!\! G(k) B_{kt_3} G_k(t_3) B_{t_3t_4} 
  \cdots B_{t_rp}G_{kt_3\cdots t_r}(p) -\tilde R_{\ell+1} \nonumber \\
  &&
\eeqn
where 
\beqn
  R_{\ell+1}&=&(-1)^\ell
   \!\!\!\!\!\!\!\!\!\!\!
    \sum_{p_2\cdots p_{\ell+1}\in\M\setminus\{k\}\atop p_i\ne p_j}
  \!\!\!\!\!\!\! \!\! 
  B_{kp_2}G_k(p_2)\cdots B_{p_{\ell-1} p_\ell} 
   \left[\{ (B_{p' p})_{ 
  p',p\in\M\setminus\{kp_2\cdots p_\ell\}}\}^{-1}\right]_{p_\ell
    p_{\ell+1}}
    B_{p_{\ell+1} k}\nonumber\\
 &&\\
 &&\nonumber\\
 \tilde R_{\ell+1}&=& (-1)^\ell \!\!\!\!\!\!\!  
 \!\!\! \sum_{t_3\cdots t_{\ell+1}\in\M\setminus\{k,p\}\atop t_i\ne t_j}
  \!\!\!\!\!\! G(k) B_{kt_3} 
  \cdots G_{kt_3\cdots t_{\ell-1}}(t_\ell)B_{t_\ell t_{\ell+1}} 
  \left[\{ (B_{p' p})_{p',p\in\M\setminus\{k t_3\cdots t_\ell\}}\}^{-1}
   \right]_{t_{\ell+1} p} \nonumber\\
 &&
\eeqn
Since $R_{N+1}=\tilde R_{N+1}=0$ the theorem follows
 $\blacksquare$ 
\goodbreak
%
%
%
%

\bigskip
\bigskip
\bigskip
\section{Application to the Many-Electron System and to 
  the $\varphi^4$-Model}

\bigskip
\bigskip
\subsection{\bf The Many-Electron System}

\bigskip
\setcounter{equation}{4}
\label{sec4}
We consider the  many-electron system in the grand canonical 
 ensemble in finite volume 
 $[0,L]^d$ and at some small but positive temperature $T=1/\beta>0$ 
 with attractive delta-interaction given by the Hamiltonian 
$$H=H_0-\lambda H_{\rm int}={\ts
{1\over L^d}}\sum_{\bk\sigma}({\ts {\bk^2\over 2m}}-\mu)a_{\bk\sigma}^+ 
 a_{\bk\sigma}-{\ts {\lambda\over L^{3d}}} \sum_{\bk\p\q} a_{\bk\up}^+ 
 a_{\q-\bk\down}^+ a_{\q-\p\down} a_{\p\up} \eqno (4.1)$$
Our normalization conventions concerning the volume factors are such 
 that the canonical anticommutation relations read $\{a_{\bk\sigma},
  a_{\p\tau}^+\}=L^d\,\delta_{\bk,\p}\delta_{\sigma,\tau}$. The  
 momentum sums range over some subset of $\left({2\pi\over L}\Bbb Z\right)^d$, 
 say ${\cal M}=\bigl\{\bk\in \left({2\pi\over L}\Bbb Z\right)^d\,\bigr|\> 
  |e_\bk|\le 1\bigr\}$, $e_\bk=\bk^2/2m-\mu$, 
  and $\q\in\{\bk-\p\,|\bk,\p\in\M\}$. 

We are interested in the momentum distribution 
$$\la a_{\bk\sigma}^+a_{\bk\sigma}\ra= Tr[e^{-\beta H}a_{\bk\sigma}^+
  a_{\bk\sigma}]\bigr/ Tr\,e^{-\beta H}  \eqno (4.2)$$
and in the expectation value of the energy 
$$\la H_{\rm int}\ra=\sum_{\q} \Lambda(\q)  \eqno (4.3)$$
where 
$$\Lambda(\q)={\ts {\lambda\over L^{3d}}}\sum_{\bk,\p} Tr[e^{-\beta H}
    a_{\bk\up}^+ a_{\q-\bk\down}^+ a_{\q-\p\down} a_{\p\up}]
  \bigr/ Tr\,e^{-\beta H} \eqno (4.4)$$

By writing down the perturbation series for the partition function, rewriting 
 it as a Grassmann integral
\beqn
 {\ts {Tr\,e^{-\beta( H_0-\lambda H_{\rm int})}
   \over Tr\, e^{-\beta H_0}}}&=&\int e^{ {\lambda\over 
  (\beta L^d)^3}\sum_{kpq}\bar\psi_{k\up}\bar\psi_{q-k\down}
    \psi_{q-p\down}\psi_{p\up} } d\mu_C(\psi,\bar\psi)\\
 d\mu_C&=&\pro_{k\sigma}{\ts {\beta L^d\over ik_0-e_\bk}}\> 
   e^{-{1\over \beta L^d}\sum_{k\sigma}(ik_0-e_\bk)\bar\psi_{k\sigma}\psi_{k\sigma}} 
   \pro_{k\sigma} d\psi_{k\sigma} d\bar\psi_{k\sigma} \>, \nonumber
\eeqn
 performing a Hubbard-Stratonovich 
 transformation ($\phi_q=u_q+iv_q$, 
   $d\phi_q d\bar\phi_q:=du_q dv_q$)
$$ e^{-\sum_q a_q b_q}= \int  e^{i\sum_q(a_q\phi_q+b_q\bar\phi_q)}  
    e^{-\sum_q|\phi_q|^2}\pro_q\ts {d\phi_q d\bar\phi_q\over \pi}
  \eqno (4.6)$$
with 
$$ a_q={\ts {\lambda^{1\over 2}\over (\beta L^d)^{3\over2}}}\sum_k
        \bar\psi_{k\up} \bar\psi_{q-k\down},\;\;\;
  b_q={\ts {\lambda^{1\over 2}\over (\beta L^d)^{3\over2}}}
   \sum_p \psi_{p\up} \psi_{q-p\down} \eqno (4.7)$$
and then integrating out the $\psi,\bar\psi$ variables, 
  one arrives at the following representation which is the 
starting point for our analysis (for more details, see [FKT] or [L1]):
$$ {\ts {1\over L^d}}\la a_{\bk\sigma}^+a_{\bk\sigma}\ra=
 {\ts {1\over \beta L^d}\>{1\over \beta}}\!\!\!\!\sum_{k_0\in {\pi\over\beta}
  (2\Bbb Z+1)} \!\!\la \psi_{\bk k_0\sigma}^+\psi_{\bk k_0\sigma}\ra 
   \eqno (4.8)$$
where, abbreviating $k=(\bk,k_0)$, $\kappa=\beta L^d$, $a_k=ik_0-e_\bk$, 
  $g=\lambda^{1\over2}$,  
$${\ts {1\over\kappa}}\,\la \bar\psi_{t\sigma}\psi_{t\sigma}\ra=\int \left[ 
 \bmatr{cc} a_{k}\delta_{k,p} & {ig\over \sqrt{\kappa}}\,
  \bar\phi_{p-k} \\ {ig\over \sqrt{\kappa}}\, \phi_{k-p} & 
  a_{-k}\delta_{k,p} \ematr \right]_{t\sigma,t\sigma}^{-1} 
 \!\!\! dP(\phi) \eqno(4.9)$$
and $dP(\phi)$ is the normalized measure 
$$dP(\phi)={\ts {1\over Z}} \>\det\left[ 
  \bmatr{cc} a_{k}\delta_{k,p} & {ig\over \sqrt{\kappa}}\,
  \bar\phi_{p-k} \\ {ig\over \sqrt{\kappa}}\, \phi_{k-p} & 
  a_{-k}\delta_{k,p} \ematr
  \right]  e^{-\sum_q|\phi_q|^2} \pro_q {\ts {d\phi_q 
 d\bar\phi_q}}  \eqno (4.10)$$
 Furthermore 
$$\Lambda(\q)={\ts {1\over\beta}}\sum_{q_0\in {2\pi\over\beta}\Bbb Z} 
  \Lambda(\q,q_0) \eqno (4.11)$$
where
\setcounter{equation}{11}
\beqn
 \Lambda(q)&=&{\ts {\lambda\over (\beta L^d)^3}}\sum_{k,p} 
  \la \bar\psi_{k\up}\bar\psi_{q-k\down}\psi_{q-p\down}\psi_{p\up}\ra
  \nonumber \\
 &=& \la |\phi_q|^2\ra -1 
\eeqn
and the expectation in the last line is integration with respect 
 to $dP(\phi)$. 
  The expectation on the 
 $\psi$ variables 
 $\la \bar\psi_{k\sigma}\psi_{k\sigma}\ra={\ts {1\over{\cal Z}}} 
 \int \bar\psi_{k\sigma}\psi_{k\sigma}\,
   e^{ {\lambda\over \kappa^3}\sum_{k,p,q}\bar\psi_{k\up}
    \bar\psi_{q-k\down}\psi_{q-p\down}\psi_{p\up} } d\mu_C$
 is Grassmann integration,  but these 
representations are not used in the following. 
 The matrix and the integral in (4.9) become finite dimensional if we
choose some cutoff on the $k_0$ variables which is removed in the
end. The set $\M$ for the spatial momenta is already finite since we 
have chosen a fixed UV-cuttoff $|e_\bk|=|\bk^2/2m-\mu|\le 1$ which
  will not be removed 
 in the end since we are interested in the infrared properties at 
 $\bk^2/2m=\mu$. 

Our goal is to apply the inversion formula to the inverse matrix element in
(4.9). Instead of writing the matrix in terms of four $N\times N$ blocks 
 $(a_k\delta_{k,p})_{k,p}$, $(\bar\phi_{p-k})_{k,p}$, 
 $(\phi_{k-p})_{k,p}$ and $(a_{-k}\delta_{k,p})_{k,p}$
 where $N$ is the number of the $d+1$-dimensional momenta $k,p$, 
 we interchange rows and columns to rewrite it in terms of $N$ blocks 
 of size $2\times 2$ (the matrix $U$ in the next line interchanges the rows and 
 columns): 
$$U\left[ 
  \bmatr{cc} a_{k}\delta_{k,p} & {ig\over \sqrt{\kappa}}\,
  \bar\phi_{p-k} \\ {ig\over \sqrt{\kappa}}\, \phi_{k-p} & 
  a_{-k}\delta_{k,p} \ematr \right]U^{-1}=B=(B_{kp})_{k,p}$$
 where the  $2\times 2$ blocks $B_{kp}$ are given by  
$$ B_{kk}=\left(
  \bmatr{cc} a_{k} & {ig\over \sqrt{\kappa}}\,
  \bar\phi_{0} \\ {ig\over \sqrt{\kappa}}\, \phi_{0} & 
  a_{-k} \ematr \right),\;\;\;B_{kp}={\ts {ig\over \sqrt{\kappa}}} 
 \left( \bmatr{cc} 0  &\bar\phi_{p-k} \\  \phi_{k-p} & 0 
    \ematr \right)\;\;{\rm if}\;k\ne p\,. \eqno (4.13)   $$
We want to compute the $2\times 2$ matrix 
$$ \la G\ra(k)=\int G(k)\,dP(\phi) \eqno (4.14)$$
where 
$$G(k)=[B^{-1}]_{kk} \eqno (4.15)$$

\bigskip
We start again with the two loop approximation which retains only the
$r=2$ term in the  denominator of (1.3). The result will be equation  
 (4.20) below where the quantities $\la\sigma_k\ra$ 
and $\la|\phi_0|^2\ra$
 appearing in (4.20) have to satisfy the equations (4.21) and (4.24) 
which have to be 
solved in conjunction with (4.29). The solution to these equations is discussed 
 below (4.30).
\par
We first derive (4.20). In the two loop approximation,
\setcounter{equation}{15} 
\beqn
  G(k)&\approx& \biggl[ B_{kk}-\sum_{p\ne k} B_{kp}\,
    G_k(p)\, B_{pk} \biggr]^{-1}  \nonumber \\
 &=&\biggl[\left(\bmatr{cc} a_{k} & {ig\over \sqrt{\kappa}}\,
  \bar\phi_{0} \\ {ig\over \sqrt{\kappa}}\, \phi_{0} & 
  a_{-k} \ematr\right) +{\ts {\lambda\over\kappa}} 
  \sum_{p\ne k} \left(\bmatr{cc}   & 
  \bar\phi_{p-k} \\ \phi_{k-p} & 
    \ematr\right) \,G_k(p)\, 
  \left(\bmatr{cc}   & 
  \bar\phi_{k-p} \\ \phi_{p-k} & 
    \ematr\right) \biggr]^{-1}\nonumber \\
 &=:& \biggl[\left(\bmatr{cc} a_{k} & {ig\over \sqrt{\kappa}}\,
  \bar\phi_{0} \\ {ig\over \sqrt{\kappa}}\, \phi_{0} & 
  \bar a_{k} \ematr\right) 
  + \Sigma(k) \biggr]^{-1}  
\eeqn
where, substituting again $G_k(p)$ by $G(p)$  in the infinite volume limit, 
$$\Sigma(k)={\ts {\lambda\over\kappa}} 
  \sum_{p\ne k} \left(\bmatr{cc}   & 
  \bar\phi_{p-k} \\ \phi_{k-p} & 
    \ematr\right)\, \biggl[ \left(\bmatr{cc}
   a_{p} & {ig\over \sqrt{\kappa}}\,
  \bar\phi_{0} \\ {ig\over \sqrt{\kappa}}\, \phi_{0} & 
  \bar a_{p} \ematr\right)+ \Sigma(p) \biggr]^{-1}\, 
  \left(\bmatr{cc}   & 
  \bar\phi_{k-p} \\ \phi_{p-k} & 
     \ematr\right)   \eqno (4.17)$$
Anticipating the fact that the off-diagonal elements of $\la\Sigma\ra(k)$ 
 will be zero (for `zero external field'), we make the Ansatz 
$$\Sigma(k)=\left(\bmatr{cc} \sigma_k& \\ & \bar\sigma_k\ematr\right)
    \eqno (4.18) $$
and obtain 
$$ \left(\bmatr{cc} \sigma_k& \\ & \bar\sigma_k\ematr\right)=
   {\ts {\lambda\over\kappa}} 
  \sum_{p\ne k} \ts {1\over (a_p+\sigma_p)(\bar a_p+\bar\sigma_p) 
   +{\lambda\over\kappa}|\phi_0|^2} 
\left(\bmatr{cc} (a_k+\sigma_k)|\phi_{p-k}|^2 
 & -{ig\over \sqrt{\kappa}}\, \phi_{0}\bar\phi_{k-p}\bar\phi_{p-k}\\
   -{ig\over \sqrt{\kappa}}\, \bar\phi_{0} \phi_{k-p}\phi_{p-k} 
  &  (\bar a_k+\bar\sigma_k)|\phi_{k-p}|^2\ematr\right)
  \eqno(4.19)$$
As for the Anderson model, we perform the functional integral by substituting 
 the quantities $|\phi_q|^2$ by their expectation values $\la|\phi_q|^2\ra$. 
 Apparently this is less obvious in this case since  $dP(\phi)$ is 
 no longer Gaussian and the $|\phi_q|^2$ are no longer identically, independently 
 distributed.  We will comment on this  
  after (4.37) below and at the end of  
 the next section by reinterpreting this procedure as a 
 resummation of diagrams. 
  For now, we simply continue in this way. Then 
$$\la G\ra(k)=\ts {1\over |a_k+\la\sigma_k\ra|^2
   +{\lambda\over\kappa}\la|\phi_0|^2\ra}\left(\bmatr{cc} \bar a_k
   +\la\bar\sigma_k\ra & \ts -{ig\over \sqrt\kappa}\,\la\bar\phi_0\ra \\ 
  -{ig\over \sqrt\kappa} \,\la\phi_0\ra  & a_k+\la\sigma_k\ra 
   \ematr\right) \eqno (4.20)$$
where the quantity $\la\sigma_k\ra$ has to satisfy the equation 
$$\la\sigma_k\ra= {\ts {\lambda\over\kappa}} 
  \sum_{p\ne k} \ts {\bar a_p+\la\bar\sigma_p\ra 
  \over |a_p+\la\sigma_p\ra|^2
   +{\lambda\over\kappa}\la |\phi_0|^2\ra }\, 
   \la |\phi_{p-k}|^2\ra \eqno (4.21) $$   

\medskip
Since $dP(\phi)$ is not Gaussian, we do not know the expectations 
 $\la|\phi_q|^2\ra$. However, by partial integration, we obtain 
$$ \la |\phi_q|^2\ra =1+ {\ts {ig\over \sqrt\kappa}} 
   \sum_p   \int \phi_q\, [B^{-1}(\phi)]_{p\up,p+q\down}\, dP(\phi) 
  \eqno(4.22) $$
Namely, 
\beq
  \la |\phi_q|^2\ra&=&{\ts {1\over Z}}\int \phi_q\bar\phi_q\, 
   \det\left[ \{B_{kp}(\phi)\}_{k,p}\right]\, e^{-\sum_q |\phi_q|^2}
      d\phi_qd\bar\phi_q \\
 &=&1+{\ts {1\over Z}}\int \phi_q\,\ts \Bigl( {\pt\over \pt\phi_q}
   \det\left[ \{B_{kp}(\phi)\}_{k,p}\right]\Bigr)\, e^{-\sum_q |\phi_q|^2}
      d\phi_qd\bar\phi_q \\
 &=&1+{\ts {1\over Z}}\int \phi_q\,\sum_{p,\tau} \ts 
   \det\left[\bmatr{ccc} |&|&|\\ 
   B_{k\sigma,p'\tau'}& \ts  {\pt B_{k\sigma,p\tau}\over \pt\phi_q}& 
   B_{k\sigma,p''\tau''} \\  |&|&|\ematr
    \right] e^{-\sum_q |\phi_q|^2}
      d\phi_qd\bar\phi_q 
\eeq
Since 
$$\ts  {\pt \over \pt\phi_q}B_{kp}
  ={ig\over \sqrt\kappa}\left(\bmatr{cc} 0&0 \cr1&0 \ematr\right)
   \,\delta_{k-p,q}$$
we have 
$$ {\det\left[\bmatr{ccc} |&|&|\\ 
   B_{k\sigma,p'\tau'}& \ts  {\pt B_{k\sigma,p\tau}\over \pt\phi_q}& 
   B_{k\sigma,p''\tau''} \\  |&|&|\ematr 
    \right]\Bigr/ \det\left[ \{B_{kp}\}_{k,p}\right]} 
  =\left\{\begin{array}{ll} 0& {\rm if}\; \tau=\down\\ & \\
  \ts {ig\over \sqrt\kappa}\, [B^{-1}]_{p\up,p+q\down}& 
 {\rm if}\; \tau=\up\end{array}\right. $$
which results in (4.22). 

The inverse matrix element in (4.22) we compute again with (1.3,4) in the 
 two loop approximation.  
Consider first the case $q=0$. Then one gets 
\setcounter{equation}{22}
\beqn
 \la |\phi_0|^2\ra&=&1+ {\ts {ig\over \sqrt\kappa}} 
   \sum_p   \int \phi_0\, G(p)_{\up\down}\, dP(\phi)  \nonumber\\
 &=&1+ {\ts {ig\over \sqrt\kappa}} 
   \sum_p   \int \phi_0\,  \ts {1\over |a_p+\sigma_p|^2
   +{\lambda\over\kappa}|\phi_0|^2}\left(\bmatr{cc}\bar a_p
   +\bar\sigma_p & \ts -{ig\over \sqrt\kappa}\,\bar\phi_0 \\ 
  -{ig\over \sqrt\kappa} \,\phi_0  & a_p+\sigma_p\ematr\right)_{\up\down}  
   \, dP(\phi)  \nonumber\\
  &=&1+ {\ts {\lambda\over \kappa}} 
   \sum_p   \int \phi_0\,  \ts {\bar\phi_0\over |a_p+\sigma_p|^2
   +{\lambda\over\kappa}|\phi_0|^2} \, dP(\phi) 
\eeqn
Performing the functional integral by substitution of expectation values
gives 
$$\la |\phi_0|^2\ra=1+ {\ts {\lambda\over \kappa}} 
   \sum_p  \la |\phi_0|^2\ra\,  \ts {1\over |a_p+\la\sigma_p\ra|^2
   +{\lambda\over\kappa}\la|\phi_0|^2\ra }  $$
or 
$$\la |\phi_0|^2\ra={1\over 1-{\ts {\lambda\over \kappa}} 
  {\ds \sum_p^{\phantom{,}}} \ts {1\over |a_p+\la\sigma_p\ra|^2
   +{\lambda\over\kappa}\la|\phi_0|^2\ra }  } \eqno (4.24) $$
Before we discuss (4.24), we write down the equation 
 for $q\ne 0$. In that case 
 we use (1.4) to compute $[B^{-1}(\phi)]_{p\up,p+q\down}$ in the two loop 
 approximation. We get 
\setcounter{equation}{24}
\beqn
\lefteqn{ [B^{-1}(\phi)]_{p\up,p+q\down}
    \approx -\left[G(p)B_{p,p+q}G(p+q)\right]_{\up\down} }\nonumber \\
 & &\nonumber \\
 \lefteqn{ =- \ts {1\over |a_p+\sigma_p|^2
   +{\lambda\over\kappa}|\phi_0|^2}
  \, \ts {1\over |a_{p+q}+\sigma_{p+q}|^2
   +{\lambda\over\kappa}|\phi_0|^2}\,{ig\over\sqrt\kappa}\times }\nonumber \\
 & & \nonumber \\    
 & & \left(\bmatr{cc} -{ig\over \sqrt\kappa}[
  (\bar a+\bar \sigma)_{p+q}\bar\phi_0\phi_{-q}+(\bar a+\bar \sigma)_{p} 
  \phi_0\bar\phi_q]& 
   (\bar a+\bar \sigma)_{p}(a+\sigma)_{p+q}\bar\phi_q 
  -{\lambda\over\kappa} \bar\phi_0^2\phi_{-q}\\
  ( a+ \sigma)_{p}(\bar a+\bar\sigma)_{p+q}\phi_{-q}
  -{\lambda\over\kappa} \phi_0^2\bar\phi_{q} & 
  -{ig\over \sqrt\kappa}[
  ( a+ \sigma)_{p+q}\phi_0\bar\phi_{q}+( a+ \sigma)_{p} 
  \bar\phi_0\phi_{-q}]  \ematr\right)_{\up\down} \nonumber \\
 & & \nonumber \\
  \lefteqn{ =- \ts{ig\over\sqrt\kappa}
   {(\bar a+\bar \sigma)_{p}(a+\sigma)_{p+q}\bar\phi_q 
  -{\lambda\over\kappa} \bar\phi_0^2\phi_{-q}
       \over \left(|a_p+\sigma_p|^2
   +{\lambda\over\kappa}|\phi_0|^2\right)
     \left( |a_{p+q}+\sigma_{p+q}|^2
   +{\lambda\over\kappa}|\phi_0|^2\right)}   } 
\eeqn
which gives

\setcounter{equation}{25}
\beqn
 \la |\phi_q|^2\ra&=&1+ {\ts {\lambda\over \kappa}} 
   \sum_p   \int \phi_q\, \ts
  {(\bar a+\bar \sigma)_{p}(a+\sigma)_{p+q}\bar\phi_q 
  -{\lambda\over\kappa} \bar\phi_0^2\phi_{-q}
       \over \left(|a_p+\sigma_p|^2
   +{\lambda\over\kappa}|\phi_0|^2\right)\left( |a_{p+q}+\sigma_{p+q}|^2
   +{\lambda\over\kappa}|\phi_0|^2\right)} \, dP(\phi)
  \nonumber  \\
 &=&1+  {\ts {\lambda\over \kappa}} 
   \sum_p   \ts
  {(\bar a_p+ \la\bar\sigma_p\ra)(a_{p+q}+\la\sigma_{p+q}\ra)
  \la |\phi_q|^2\ra -{\lambda\over\kappa} \la\bar\phi_0^2
   \phi_q\phi_{-q}\ra 
       \over \left(|a_p+\la\sigma_p\ra|^2
   +{\lambda\over\kappa}\la|\phi_0|^2\ra\right)
  \left( |a_{p+q}+\la\sigma_{p+q}\ra|^2
   +{\lambda\over\kappa}\la|\phi_0|^2\ra \right)}   
\eeqn
Although one may think that the expectation $\la\bar\phi_0^2
   \phi_q\phi_{-q}\ra $ vanishes for zero external field, this is not so. 
This can be seen again by partial integration:
\beq
  \la\bar\phi_0^2
   \phi_q\phi_{-q}\ra &=&{\ts {1\over Z}}\int\bar\phi_0^2
   \phi_q\phi_{-q}\, 
   \det\left[ \{B_{kp}(\phi)\}_{k,p}\right]\, e^{-\sum_q |\phi_q|^2}
      d\phi_qd\bar\phi_q \\
 &=&{\ts {1\over Z}}\int \bar\phi_0^2\phi_q\,\ts 
   \Bigl( {\pt\over \pt\bar\phi_{-q}}
   \det\left[ \{B_{kp}(\phi)\}_{k,p}\right]\Bigr)\, e^{-\sum_q |\phi_q|^2}
      d\phi_qd\bar\phi_q \\
 &=&{\ts {1\over Z}}\int \bar\phi_0^2\phi_q\,\sum_{p,\tau} \ts 
   \det\left[\bmatr{ccc} |&|&|\\ 
   B_{k\sigma,p'\tau'}& \ts  {\pt B_{k\sigma,p\tau}\over
     \pt\bar\phi_{-q}}& 
   B_{k\sigma,p''\tau''} \\  |&|&|\ematr
    \right] e^{-\sum_q |\phi_q|^2}
      d\phi_qd\bar\phi_q 
\eeq
The above determinant is multiplied and devided by 
 $\det\left[ \{B_{kp}\}_{k,p}\right] $ to give 
$$ {\det\left[\bmatr{ccc} |&|&|\\ 
   B_{k\sigma,p'\tau'}& \ts  {\pt B_{k\sigma,p\tau}\over 
  \pt\bar\phi_{-q}}& 
   B_{k\sigma,p''\tau''} \\  |&|&|\ematr 
    \right]\Bigr/ \det\left[ \{B_{kp}\}_{k,p}\right]} 
  =\left\{\begin{array}{ll} 0& {\rm if}\; \tau=\up\\ & \\
  \ts {ig\over \sqrt\kappa}\, [B^{-1}]_{p\down,p+q\up}& 
 {\rm if}\; \tau=\down\end{array}\right. $$
Computing the inverse matrix element again in the two loop 
approximation (4.25), we arrive at 
$$ \la\bar\phi_0^2 \phi_q\phi_{-q}\ra =
{\ts {\lambda\over\kappa}} \sum_p \Bigl\la \ts
  { (a_p+\sigma_p)(\bar a_{p+q}+\sigma_{p+q})
  \bar\phi_0^2\phi_q\phi_{-q} -{\lambda\over\kappa}\bar\phi_0^2
  \phi_0^2 \phi_q\bar\phi_q \over \left(|a_p+\sigma_p|^2
   +{\lambda\over\kappa}|\phi_0|^2\right)
     \left( |a_{p+q}+\sigma_{p+q}|^2
   +{\lambda\over\kappa}|\phi_0|^2\right)} \Bigr\ra $$
Abbreviating 
$$ g_p=\ts { a_p+\la \sigma_p\ra \over |a_p+\la\sigma_p\ra|^2
   +{\lambda\over\kappa}\la|\phi_0|^2\ra},\;\;\;\;\;\;
  f_p=\ts {\sqrt{ {\lambda\over\kappa}\la|\phi_0|^2\ra}
  \over |a_p+\la\sigma_p\ra|^2
   +{\lambda\over\kappa}\la|\phi_0|^2\ra} \eqno (4.27)$$
this gives 
$${\ts {\lambda\over\kappa}} \la\bar\phi_0^2 \phi_q\phi_{-q}\ra =
  {\ts {\lambda\over\kappa}} \sum_p g_p \bar g_{p+q}\, 
 {\ts {\lambda\over\kappa}}  \la\bar\phi_0^2 \phi_q\phi_{-q}\ra 
  -  {\ts {\lambda\over\kappa}} \sum_p f_p f_{p+q}\,
  {\ts{\lambda\over\kappa}}\la|\phi_0|^2\ra
    \,\la |\phi_q|^2\ra$$
or 
$${\ts {\lambda\over\kappa}} \la\bar\phi_0^2 \phi_q\phi_{-q}\ra =
  {-{\lambda\over\kappa}\sum_p f_pf_{p+q}\,
  {\lambda\over\kappa}\la|\phi_0|^2\ra
  \over 1-{\lambda\over\kappa}\sum_p g_p \bar g_{p+q}  } 
  \, \la |\phi_q|^2\ra 
  \eqno (4.28)$$
Substituting this in (4.26), we finally arrive at 
$$ \la |\phi_q|^2\ra ={ 1-{\lambda\over\kappa}\sum_p g_p 
   \bar g_{p+q} \over \bigl|1-{\lambda\over\kappa}\sum_p g_p 
   \bar g_{p+q} \bigr|^2-\bigl( {\lambda\over\kappa}\sum_p 
   f_p f_{p+q} \bigr)^2 }  \eqno(4.29)$$
where $g_p,f_p$ are given by (4.27). 
Observe that, since $dP(\phi)$ is complex, also $\la |\phi_q|^2\ra $ 
is in general complex. Only after summation over the $q_0$ variables 
we obtain necessarily a real quantity which is given by (4.4,11). 

\bigskip
We now discuss the solutions to (4.24) and (4.29). We assume that 
the solution 
 $\la \sigma_k\ra$ of (4.21) is sufficiently small such that the 
 BCS equation 
$$  {\ts {\lambda\over \kappa}} 
   \sum_p^{\phantom{,}} { \ts {1\over |a_p+\la\sigma_p\ra|^2
   + |\Delta|^2  }} = 1 \eqno (4.30)$$
has a nonzero solution $\Delta\ne 0$ (in particular this excludes 
large corrections like $\la \sigma_p\ra\sim p_0^\alpha$, 
 $\alpha\le 1/2$, which one may expect in the case of 
 Luttinger  liquid behaviour, for $d=1$ one should make 
a seperate analysis),  and make the Ansatz 
$$\lambda\la|\phi_0|^2\ra= 
  \beta L^d\, |\Delta|^2 +\eta \eqno (4.31) $$
where $\eta$ is independent of the volume. Then 
\setcounter{equation}{31}
\beqn
  {\ts {\lambda\over \kappa}} 
 \sum_p^{\phantom{,}} \ts {1\over |a_p+\la\sigma_p\ra|^2
   +{\lambda\over\kappa}\la|\phi_0|^2\ra } &= &
  {\ts {\lambda\over \kappa}} 
 \sum_p^{\phantom{,}} \ts {1\over |a_p+\la\sigma_p\ra|^2
   + |\Delta|^2+{\eta\over\kappa} }  \nonumber\\
 &=&{\ts {\lambda\over \kappa}} 
 \sum_p^{\phantom{,}} {\ts {1\over |a_p+\la\sigma_p\ra|^2
   + |\Delta|^2 }} -
  {\ts {\lambda\over \kappa}} 
 \sum_p^{\phantom{,}} \ts {\eta/\kappa\over ( |a_p+\la\sigma_p\ra|^2
   + |\Delta|^2)^2 } +O\left(({\eta\over\kappa})^2\right) \nonumber\\
 &=&\ts 1 - c_\Delta\,{\eta\over\kappa}+O\left(({\eta\over\kappa})^2\right) 
\eeqn
where we put $c_\Delta=
  {\ts {\lambda\over \kappa}} 
 \sum_p \ts {1\over ( |a_p+\la\sigma_p\ra|^2
   + |\Delta|^2)^2 }$ and used the BCS equation (4.30) in the last line. 
Equation (4.24) becomes 
\beq
  \kappa\, |\Delta|^2 +\eta &=&{\lambda\over 
    c_\Delta \,{\eta\over\kappa} 
      +O\left(({\eta\over\kappa})^2\right) } 
  =\kappa\,{\lambda\over c_\Delta\,\eta} +O(1) 
\eeq
 and has a solution $\eta=\lambda/(c_\Delta|\Delta|^2)$.  

\par
Now consider 
 $\la|\phi_q|^2\ra$ for small but nonzero $q$. In the limit $q\to 0$ 
the denominator in (4.29) vanishes, or more precisely, is of order 
 $O(1/\kappa)$ since 
$$1-{\ts {\lambda\over\kappa}}\sum_p g_p 
   \bar g_{p} -{\ts {\lambda\over\kappa}}\sum_p 
   f_p f_{p}\; =\;1-{\ts {\lambda\over\kappa}}\sum_p\ts {1\over 
   |a_p+\la \sigma_p\ra|^2+{\lambda\over\kappa} 
  \la |\phi_0|^2\ra}\;=\;O(1/\kappa)$$
because  of (4.32). If we assume the second derivatives of 
 $\la\sigma_k\ra$ to be integrable (which should be the case 
for $d=3$ and $\la|\phi_q|^2\ra\sim 1/q^2$ by virtue of (4.21)), 
then, since the denominator in (4.29) is an even function of $q$, the 
small $q$ behaviour of $\la|\phi_q|^2\ra$ is $1/q^2$. 
 This agrees with the common 
expectations [FMRT,CFS,B]. Usually the behaviour of
$\la|\phi_q|^2\ra$  is infered from 
the second order Taylor expansion of the effective potential
$$V_{\rm eff}(\{\phi_q\})=\sum_q|\phi_q|^2-\log
  \det\left[ 
  \bmatr{cc} \delta_{k,p} & {ig\over \sqrt{\kappa}}\,
  {\bar\phi_{p-k}\over a_k} \\
  {ig\over \sqrt{\kappa}}\, {\phi_{k-p}\over a_{-k}} & 
  \delta_{k,p} \ematr
  \right]   \eqno (4.33) $$
around its global minimum [L2]
$$\phi_q^{\rm min}=\sqrt{\beta L^d}\,\ts {|\Delta|\over \sqrt\lambda}\,\delta_{q,0}\, 
  e^{i\theta_0}  \eqno (4.34) $$
where the phase $\theta_0$ of $\phi_0$ is arbitrary. If one expands $V_{\rm eff}$ up to 
 second order in 
$$ \xi_{q}=\phi_{q}-\delta_{q,0}\sqrt{\beta L^d}\,
   \ts {|\Delta|\over \sqrt\lambda}\,
    e^{i\theta_0}  
  =\left\{ \begin{array}{ll}
    \left(\rho_0-\sqrt{\beta L^d}\,\ts {|\Delta|\over \sqrt\lambda}\right)
   \,e^{i\theta_0} &\mbox{for $q=0$ }\\ 
   \>\rho_{q}\, e^{i\theta_{q}} &\mbox{for $q\ne 0$} 
 \end{array}\right. \eqno(4.35) $$
one obtains [L2] 
\setcounter{equation}{35}
\beqn
  V_{\rm eff}(\{\phi_q\})&=&V_{\rm min}+2\beta_0 \,
  (\rho_0-\sqrt{\beta L^d}\,{\ts {|\Delta|\over \sqrt\lambda}})^2 +
  \sum_{q\ne 0} (\alpha_q+i\gamma_q)\, \rho_q^2 \nonumber\\
 &\phantom{m}& + {\ts{1\over2}}\sum_{q\ne 0} 
   \beta_q \,|e^{-i\theta_0} \phi_q
   +e^{i\theta_0} \bar\phi_{-q}|^2 
  +O(\xi^3)  
\eeqn
where for small $q$ one has $\alpha_q,\gamma_q\sim q^2$. 
Hence, if $V_{\rm eff}$ is substituted by the right hand side of (4.36) 
 one obtains 
 $\la |\phi_q|^2\ra\sim 1/q^2$. 

\par
For $d=3$, this seems to be the right answer, but in lower dimensions 
one would expect an integrable singularity due to 
 (4.21) and (4.3,4,11). In particular, we think it would be a very 
interesting problem to solve 
the integral equations (4.21,24,29) for $d=1$ and to check the result for 
Luttinger liquid behaviour. A good warm up excercise would be to
consider the $0+1$ dimensional problem, that is, we only have the 
$k_0,p_0,q_0$-variables. In that case the `bare BCS equation' 
$${\ts {\lambda\over\beta}}
 \sum_{p_0\in {\pi\over\beta}(2\Bbb Z+1)} \ts {1\over p_0^2+ 
  |\Delta|^2} =1 $$
still has a nonzero solution $\Delta$ for sufficiently small 
 $T=1/\beta$ and the question would be whether the 
correction $\la\sigma_{p_0}\ra$ is sufficiently big to destroy the gap. 
 That is, does the `renormalized BCS equation' 
$${\ts {\lambda\over\beta}}
 \sum_{p_0\in {\pi\over\beta}(2\Bbb Z+1)} 
  \ts {1\over |p_0+\la\sigma_{p_0}\ra|^2+ 
  |\Delta|^2} =1 $$ 
$\la\sigma_{p_0}\ra$ being the solution to (4.21,24,29), still have a nonzero
solution? We remark that, if the gap vanishes (for arbitrary dimension), 
then also the singularity 
of $\la |\phi_q|^2\ra$ disappears. Namely, if the gap equation 
has no solution, that is, if ${1\over\kappa}\sum_p {1\over |a_p+
  \la \sigma_p\ra|^2}<\infty$, then $\la |\phi_0|^2\ra$ 
given by (4.24) is no longer macroscopic (for sufficiently small coupling) 
and ${\lambda\over\kappa}\la |\phi_0|^2\ra$ vanishes in the 
infinite volume limit. And the denominator in (4.29) becomes for 
 $q\to 0$ 
$$1-{\ts {\lambda\over\kappa}}\sum_p \ts {1\over 
  |a_p+\la \sigma_p\ra|^2}$$
which would be nonzero (for sufficiently small coupling). 

\bigskip
Finally we argue why it is reasonable to substitute $|\phi_0|^2$ by its 
expectation
value while performing the functional integral. 
 We may write the effective potential (4.33) 
as 
$$V_{\rm eff}(\{\phi_q\})=V_1(\phi_0)+ V_2(\{\phi_q\}) \eqno (4.37)$$
where 
\setcounter{equation}{37}
\beqn
  V_1(\phi_0)&=&|\phi_0|^2- \sum_{k}\ts
 \log\left[ 1+ {\lambda\over\kappa}
  { |\phi_0|^2 \over k_0^2+e_\bk^2}\right] \nonumber \\
  &=& \kappa \Bigl({\ts {|\phi_0|^2\over\kappa}- {1\over\kappa}}\sum_{k}\ts
 \log\left[ 1+ {\lambda {|\phi_0|^2\over\kappa}\over k_0^2+e_\bk^2}\right]
   \Bigr)\equiv \kappa\, 
 V_{\rm BCS}\ts\left({|\phi_0|\over\sqrt\kappa}\right)
\eeqn
and
$$ V_2(\{\phi_q\})=  \sum_{q\ne 0}|\phi_q|^2-\log
  \det\left[ 
  \left(\bmatr{cc} \delta_{k,p} & {ig\over \sqrt{\kappa}}\,
  {\bar\phi_{0}\over a_k}\delta_{k,p} \\
   {ig\over \sqrt{\kappa}}\, {\phi_{0}\over a_{-k}}\delta_{k,p} & 
  \delta_{k,p} \ematr\right)^{-1}
  \left(\bmatr{cc} \delta_{k,p} & {ig\over \sqrt{\kappa}}\,
  {\bar\phi_{p-k}\over a_k} \\
  {ig\over \sqrt{\kappa}}\, {\phi_{k-p}\over a_{-k}} & 
  \delta_{k,p} \ematr\right)
  \right]  \eqno(4.39)$$
If we ignore the $\phi_0$-dependence of $V_2$, then the $\phi_0$-integral
\setcounter{equation}{39} 
\beqn
  {\int F\left({1\over\kappa}|\phi_0|^2\right)\,
   e^{-V_{1}(\phi_0)} d\phi_0
  d\bar\phi_0 \over  
  \int e^{-V_{1}(\phi_0)} d\phi_0
  d\bar\phi_0} 
  &=& {\int F\left( \rho^2\right)\, e^{-\kappa V_{\rm BCS}(\rho)} 
   \rho\,d\rho \over  
  \int e^{-\kappa V_{\rm BCS}(\rho)}  \rho\,d\rho} 
   \buildrel \kappa\to\infty\over \to F(\rho_{\rm min}^2)
  =\ts F\left({1\over\kappa}\la|\phi_0|^2\ra\right)\nonumber \\
 & & 
\eeqn
simply puts $|\phi_0|^2$ at the global minimum 
  of the (BCS) effective potential. 
\bigskip
\bigskip \goodbreak
%
%
%
%
\subsection{The $\vp^4$-Model}
\bigskip
In this section we choose the  $\varphi^4$-model as a typical
bosonic model to demonstrate our method. As in section 2, we start in 
 finite volume $[0,L]^d$ on a lattice with lattice spacing $1/M$. 
The two point function is given by 
$$S(x,y)=\la \vph_x\vph_y\ra:={ \int_{\Bbb R^{N^d}} 
     \vph_x\vph_y\> e^{-{g^2\over 2}{1\over M^d} 
  \sum_x \vph_x^4}\> e^{-{1\over M^{2d}}\sum_{x,y} 
  (-\Delta+m^2)_{x,y}\vph_x\vph_y} \pro_x d\vph_x \over 
   \int_{\Bbb R^{N^d}} e^{-{g^2\over 2}{1\over M^d} 
  \sum_x \vph_x^4}\> e^{-{1\over M^{2d}}\sum_{x,y} 
  (-\Delta+m^2)_{x,y}\vph_x\vph_y} \pro_x d\vph_x} \eqno (4.41)$$
where 
$$(-\Delta+m^2)_{x,y}=M^d\Bigl[
  -M^2\sum_{i=1}^d(\delta_{x,y-e_i/M} +\delta_{x,y+e_i/M} 
  -2\delta_{x,y} ) +m^2\delta_{x,y}\Bigr]\,  \eqno (4.42) $$
First we have to bring this into the form $\int[P+Q]^{-1}_{x,y}d\mu$, 
 $P$ diagonal in momentum space, $Q$ diagonal in coordinate space.  
This is done again by making a Hubbard Stratonovich  transformation 
 which in this case reads 
$$  e^{-{1\over 2} \sum_x a_x^2}=\int e^{i\sum_x a_x u_x} 
  e^{-{1\over 2} \sum_x u_x^2}\pro_x  
    {\ts {d u_x\over\sqrt{2 \pi}}} \eqno (4.43)$$
with 
$$a_x={\ts   {g\over \sqrt{M^d}}}\vph_x^2  \eqno (4.44)  $$
The result is Gaussian in the $\vp_x$-variables and the integral over 
 these variables gives 
$$S(x,y)= \int_{\Bbb R^{N^d}}\ts \left[ {1\over M^{2d}} 
  (-\Delta+m^2)_{x,y}-{ig\over \sqrt{M^d}} u_x\delta_{x,y} 
   \right]_{x,y}^{-1} dP(u)  \eqno (4.45)$$
where 
$$ dP(u)={\ts{1\over  Z}} \,
  {\ts\det\left[ {1\over M^{2d}} 
  (-\Delta+m^2)_{x,y}-{ig\over \sqrt{M^d}} u_x\delta_{x,y} 
   \right]^{-{1\over2}} } e^{-{1\over 2}\sum_x u_x^2}\pro_x
    du_x \eqno  (4.46)   $$
Since we have  bosons, the determinant comes with a power of 
 $-1/2$ which is the only difference compared to a fermionic system. 
 In momentum space this reads (compare equations (2.7-11)) 
$$S(x-y)={\ts {1\over L^d}}\sum_k e^{ik(x-y)} \la G\ra (k) \eqno (4.47)$$
where ($\gamma_q=v_q+iw_q$, $\gamma_{-q}=\bar\gamma_q$, 
   $d\gamma_qd\bar\gamma_q:=dv_qdw_q$)
$$\la G\ra (k)= \int_{\Bbb R^{N^d}}\ts \left[
   a_k\delta_{k,p} -{ig\over \sqrt{L^d}}
   {\gamma_{k-p}} 
    \right]^{-1}_{kk}  dP(\gamma) \eqno (4.48)$$ 
and 
$$dP(\gamma)={\ts {1\over Z}}\, {\ts \det\left[
   a_k\delta_{k,p} -{ig\over \sqrt{L^d}}
   {\gamma_{k-p}} 
    \right]^{-{1\over 2}}} e^{-{1\over 2}v_0^2}dv_0 \pro_{q\in \M^+} 
   e^{-|\gamma_q|^2} {\ts {d\gamma_q 
  d\bar\gamma_q}} \eqno (4.49)$$
and $\M^+$ again is a set such that either $q\in \M^+$ or 
  $-q\in \M^+$. Furthermore 
$$a_k= 4M^2\sum_{i=1}^d\ts  \sin^2\left[ {k_i\over 2M}\right] \;+\; 
  m^2  \eqno (4.50)$$

\medskip
Equation (4.48) is our starting point. We apply (1.3) to the inverse matrix 
 element in (4.48). In the two loop 
 approximation one obtains $(\gamma_0=v_0\in\Bbb R)$
$${\ts \left[a_k\delta_{k,p} -{ig\over \sqrt{L^d}}
   {\gamma_{k-p}} 
    \right]^{-1}_{kk} }\approx {1\over a_k
  -{igv_0\over \sqrt{L^d}} +{g^2\over L^d}\sum_{p\ne k}G_k(p)|\gamma_{k-p}|^2 }
  =: {1\over a_k+\sigma_k}  \eqno (4.51)$$
where 
$$\sigma_k =-{\ts {ig\over \sqrt{L^d}}}v_0+
  {\ts {g^2\over L^d}}\sum_{p\ne k}{ |\gamma_{k-p}|^2
  \over a_p-{igv_0\over \sqrt{L^d}} +\sigma_p} \eqno (4.52) $$
which results in 
$$\la G\ra(k)= {1\over a_k +\la\sigma_k\ra } \eqno (4.53)$$
where $\la\sigma_k\ra$ has to satisfy the equation 
\setcounter{equation}{53}
\beqn
 \la\sigma_k\ra & =&
   -{\ts {ig\over \sqrt{L^d}}}\la v_0\ra +{\ts {g^2\over L^d}}
   \sum_{p\ne k}{ \la |\gamma_{k-p}|^2\ra
  \over a_p +\la\sigma_p\ra }\nonumber  \\
 &=& {\ts {g^2\over 2 L^d}}\sum_{p} \la G\ra(p) +{\ts {g^2\over L^d}}
   \sum_{p\ne k}{ \la |\gamma_{k-p}|^2\ra
  \over a_p +\la\sigma_p\ra } 
 ={\ts {g^2\over L^d}}
   \sum_{p\ne k}{ \la |\gamma_{k-p}|^2\ra +{1\over 2}
  \over a_p +\la\sigma_p\ra } 
\eeqn
where the last line is due to 
\beqn
  \la v_0\ra&=&{\ts {1\over Z}}\int v_0\> 
  {\ts \det\left[
   a_k\delta_{k,p} -{ig\over \sqrt{L^d}}
   {\gamma_{k-p}} 
    \right]^{-{1\over 2}}} e^{-{1\over 2}v_0^2}dv_0 \pro_{q\in \M^+} 
   e^{-|\gamma_q|^2} {\ts {d\gamma_q 
  d\bar\gamma_q}}   \nonumber \\ 
 &=&{\ts {1\over Z}}\int \Bigl\{ {\ts {\pt\over \pt v_0}}
   {\ts \det\left[
   a_k\delta_{k,p} -{ig\over \sqrt{L^d}}
   {\gamma_{k-p}} 
    \right]^{-{1\over 2}}}\Bigr\} e^{-{1\over 2}v_0^2}dv_0 \pro_{q\in \M^+} 
   e^{-|\gamma_q|^2} {\ts {d\gamma_q 
  d\bar\gamma_q}}   \nonumber\\ 
 &=&{\ts -{1\over 2}}\sum_p {\ts (-{ig\over \sqrt{L^d}}) }\int \ts
   \left[
   a_k\delta_{k,p} -{ig\over \sqrt{L^d}}
   {\gamma_{k-p}} 
    \right]^{-1}_{pp} dP(\gamma) 
\eeqn

\smallskip
As for the Many-Electron system, we can derive an equation 
 for $\la|\gamma_q|^2\ra$ by partial integration:
\beqn
  \la|\gamma_q|^2\ra&=&{\ts {1\over Z}} \int 
  \gamma_q \bar\gamma_q\>  {\ts \det\left[
   a_k\delta_{k,p} -{ig\over \sqrt{L^d}}
   {\gamma_{k-p} } 
    \right]^{-{1\over 2}}} e^{-{v_0^2\over 2}}dv_0 \pro_{q} 
   e^{-|\gamma_q|^2} {\ts {d\gamma_q 
  d\bar\gamma_q}}   \nonumber \\ 
 &=&\;1\;+\; {\ts {1\over Z}} \int 
  \gamma_q {\ts{\partial\over \partial \gamma_q}}\biggl\{
    {\ts \det\left[
  a_k \delta_{k,p} -{ig\over \sqrt{L^d}}
   {\gamma_{k-p}} 
    \right]^{-{1\over 2}}}  \biggr\}
   e^{-{v_0^2\over 2}}dv_0 \pro_{q} 
   e^{-{1\over 2}|\gamma_q|^2} {\ts {d\gamma_q 
  d\bar\gamma_q}}  \nonumber\\  
 &=&\;1\;-\;{\ts{1\over2}}\>  \int 
  \gamma_q\,{ {\partial\over \partial \gamma_q}
    {\ts \det\left[
   a_k \delta_{k,p} -{ig\over \sqrt{L^d}}
   {\gamma_{k-p}}  \right]} \over 
 {\ts \det\left[
   a_k\delta_{k,p} -{ig\over \sqrt{L^d}}
   {\gamma_{k-p}} \right] }  }  \,dP(\gamma) \nonumber\\  
 &=&\;1\;-\;{\ts{1\over2}}\>\sum_p {\ts {-ig\over \sqrt{L^d}}}
   \int \ts \gamma_q\, \left[ a_k\delta_{k,p} -{ig\over \sqrt{L^d}}
   {\gamma_{k-p}}  \right]_{p,p+q}^{-1}
    \,dP(\gamma) 
\eeqn
Computing the inverse matrix element in (4.56) again in the two loop 
approximation, one arrives at 
$$\la |\gamma_q|^2\ra=1-\la |\gamma_q|^2\ra 
  {\ts {g^2\over 2 L^d}} \sum_p \ts {1\over 
   (a_p+\la\sigma_p\ra)(a_{p+q}+\la\sigma_{p+q}\ra)}$$
or  
$$\la |\gamma_q|^2\ra={1\over 1+
   {\ts {g^2\over 2}} \int_{[0,2\pi M]^d} {d^dp\over (2\pi)^d}\, {1\over 
   (a_p+\la\sigma_p\ra)(a_{p+q}+\la\sigma_{p+q}\ra)}  }  \eqno (4.57)$$
which has to be solved in conjunction with 
$$\la\sigma_k\ra  =\ts {g^2}
   \int_{[0,2\pi M]^d}\ts   {d^dp\over (2\pi)^d}\, 
   { \la |\gamma_{k-p}|^2\ra+{1\over2} \over a_p +\la\sigma_p\ra }
 \eqno (4.58) $$
Introducing the rescaled quantities 
$$\la \sigma_k\ra= M^2 s_{p\over M}\,,\;\;\;\;
  \la |\gamma_q|^2\ra=\lambda_{q\over M}\,,\;\;\;\;
   a_k=M^2 \vep_{k\over M}\,,\;\;\vep_k=\sum_{i=1}^d \sin^2\ts{k_i\over2}
  +{m^2\over M^2}\eqno (4.60)$$
 (4.57,58) read 
\setcounter{equation}{60}
\beqn
   s_k & =&\ts M^{d-4} {g^2}
   \int_{[0,2\pi ]^d}\ts   {d^dp\over (2\pi)^d}\, 
   { \lambda_{k-p}+{1\over2} \over \vep_p + s_p}  \\
 \lambda_q&=&{1\over 1+
   {\ts M^{d-4}{g^2\over 2}} \int_{[0,2\pi ]^d} {d^dp\over (2\pi)^d}\,
   {1\over 
   (\vep_p+ s_p)(\vep_{p+q}+s_{p+q})}  } 
\eeqn
Unfortunately we cannot check this result with the rigorously proven 
 triviality theorem since 
$\la\sigma_k\ra$ and $\la|\gamma_q|^2\ra$ only 
 give information on the 2-point function $S(x,y)$, (4.41), and 
 on ${g^2\over M^d}\sum_x \la \vp(x)^4\ra=\sum_q\Lambda(q)$ where 
 $\Lambda(q)=\la|\gamma_q|^2\ra-1$. However, the triviality theorem [F,FFS] 
 makes a statement on the connected 4-point function 
 $S_{4,c}(x_1,x_2,x_3,x_4)$ at noncoinciding 
 arguments, namely that this function vanishes in the continuum limit 
 in dimension $d>4$. 

\bigskip
Before we  include the higher loop terms of (1.3,4) and give an interpretation in terms 
 of diagrams, we would like to comment shortly on a problem which
was suggested to us by A. Sokal after a preprint of this paper was
published on the web. It refers to the $\phi^2\psi^2$-  or 
$\phi_1^2\phi_2^2$-model. That is, we have two scalar bosonic 
fields on a lattice with unit lattice spacing with action 
$${\cal S}(\phi_1,\phi_2)=\sum_{i=1}^2
\sum_x\phi_i(x)(-\Delta+m^2)\phi_i(x) 
   +\lambda\sum_x \phi_1(x)^2 \phi_2(x)^2$$
The question is whether there is exponential decay (or a gap in
momentum space) for the two point function 
$G(x,y)=\int \phi_1(x)\phi_1(y)\, e^{-{\cal S}(\phi)}
  \bigr/\int e^{-{\cal S}(\phi)}$ in the zero mass $m\to 0$ limit. A 
computation with the above formalism in two loop approximation 
gives $G(k)={1\over k^2+\sigma}$ where the gap $\sigma$ has to 
satisfy the equation $\sigma=\lambda\int_{[-\pi,\pi]^d}{d^dp\over 
  (2\pi)^d}\, {1\over p^2+\sigma}$ which gives 
$$\ts\sigma=\left\{\begin{array}{rr} O(\lambda)&{\rm if}\; d\ge 3 \cr 
          O\bigl(\lambda\log[1/\lambda]\bigr)&{\rm if}\;  d=2 \cr
          O(\lambda^{2\over 3})&{\rm if}\; d=1 \end{array}
   \right. $$

\bigskip
We now include the higher loop terms of (1.3,4) and give an interpretation in terms 
 of diagrams. 
The exact equations for $\la G\ra(k)$ and $\la |\gamma_q|^2\ra$ are 
\setcounter{equation}{62}
\beqn
  \la G\ra (k)&=& \int \ts \left[
   a_k\delta_{k,p} -{ig\over \sqrt{L^d}}
   {\gamma_{k-p} }  \right]^{-1}_{kk}  dP(\gamma)
  \;=\;\left\la {1\over a_k+\sigma_k}
    \right\ra  \\
 \sigma_k&=& -{\ts {ig\over \sqrt{L^d}}}\, v_0 +\sum_{r=2}^{N^d}  
  {\ts \left({ig\over \sqrt{L^d}}\right)^r}\!\!\!\!
  \sum_{p_2\cdots p_r\ne k\atop p_i\ne p_j} 
  G_k(p_2)\cdots G_{kp_2\cdots p_{r-1}}(p_r)\,\gamma_{k-p_2}\gamma_{p_2-p_3}\cdots 
  \gamma_{p_r-k} \nonumber 
\eeqn
and 
\beqn
  \la |\gamma_q|^2\ra&=&1+{\ts {ig\over 2\sqrt{L^d}}}\sum_p 
   \int \ts \gamma_q \left[
   a_k\delta_{k,p} -{ig\over \sqrt{L^d}}
   {\gamma_{k-p} }  \right]^{-1}_{p,p+q} dP(\gamma) \\
 &\buildrel p\to p_2\over=& 1+{\ts {1\over 2}}
   \sum_{r=2}^{N^d} \left( {\ts {ig\over\sqrt{L^d}}}\right)^r 
   \!\!\!\!\!\sum_{p_2\cdots p_r\ne p_2+q\atop p_i\ne p_j}  \Bigl\la 
  G(p_2)G_{p_2}(p_3) 
   \cdots G_{p_2\cdots p_{r-1}}(p_r) G_{p_2\cdots p_r}(p_2+q)\times 
  \nonumber \\ 
 & &\phantom{mmmmmmmmmmmmmmmm } 
 \gamma_{p_2-p_3}\cdots \gamma_{p_{r-1}-p_r}\gamma_{p_r-p_2-q}\gamma_{p_2+q-p_2}
  \Bigr\ra  \nonumber
\eeqn
For $r>2$, we obtain terms $\la \gamma_{k_1}\cdots\gamma_{k_r}\ra$ 
whose connected contributions 

\begin{figure}[thb]
 \centerline{\epsfbox{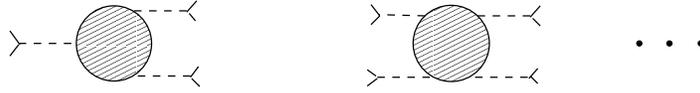}}
 \caption{three and higher loop contributions}
 \label{Figure 4}
\end{figure} 
\noindent 
are, in terms of the electron or $\vp^4$-lines, are at least six-legged. 
Since for the many-electron system and for the $\vp^4$-model (for $d=4$) 
 the relevant  diagrams are two- and four-legged [FT,R], one may start with an
approximation which ignores the connected $r$-loop contributions for $r>2$. 
This is obtained by  writing 
$$\bigl\la\gamma_{k_1}\cdots \gamma_{k_{n}} \bigr\ra \approx 
  \bigl\la\gamma_{k_1}\cdots \gamma_{k_{n}} \bigr\ra_2 \eqno (4.65)$$
where (the index `2' for `retaining only two-loop contributions')
$$\bigl\la\gamma_{k_1}\cdots \gamma_{k_{2n}} \bigr\ra_2
  :=\sum_{{\rm pairings}\;\sigma} \la\gamma_{k_{\sigma 1}}
   \gamma_{k_{\sigma 2}} \ra\cdots \la\gamma_{k_{\sigma (2n-1)}}
   \gamma_{k_{\sigma 2n}} \ra  =\int \gamma_{k_1}\cdots \gamma_{k_{2n}} \,
  dP_2(\gamma) \eqno (4.66) $$
if we define 
$$dP_2(\gamma):= e^{-\sum_q{|\gamma_q|^2\over \la |\gamma_q|^2\ra} } 
  \pro_q \ts {d\gamma_q d\bar\gamma_q\over \pi\, \la |\gamma_q|^2\ra } 
  \eqno (4.67)$$
Substituting $dP$ by $dP_2$ in (4.63,64), we obtain a model which differs 
 from the original model only by irrelevant contributions and for which we
are able to write down a closed set of equations for the two-legged 
 particle correlation function $\la G\ra (k)$ and the two-legged squiggle
 correlation function $\la |\gamma_q|^2\ra$ by resumming all two-legged 
 (particle and squiggle) subdiagrams. The exact equations of this model are
\setcounter{equation}{67} 
\beqn
  \la G\ra (k)&=& \int \ts \left[
   a_k\delta_{k,p} -{ig\over \sqrt{L^d}}
   {\gamma_{k-p} }  \right]^{-1}_{kk}  dP_2(\gamma) \\
 \la |\gamma_q|^2\ra&=&1+{\ts {ig\over 2\sqrt{L^d}}}\sum_p 
   \int \ts \gamma_q \left[
   a_k\delta_{k,p} -{ig\over \sqrt{L^d}}
   {\gamma_{k-p} }  \right]^{-1}_{p,p+q} dP_2(\gamma)
\eeqn
and  the resummation 
 of the two-legged particle and squiggle subdiagrams is 
 obtained by applying the inversion formula (1.3,4) to the 
 inverse matrix elements in (4.68,69). A discussion similar to those of  
 section 2 gives the following closed set of equations for the 
 quantities $\la G\ra(k)$ and $\la |\gamma_q|^2\ra$: 
$$\la G\ra(k)= {1\over a_k+\la \sigma_k\ra}\,,\;\;\;\;\;\;
  \la |\gamma_q|^2\ra={1\over 1+\la\pi_q\ra}  \eqno (4.70)$$
where 
\setcounter{equation}{70} 
\beqn
 \la \sigma_k\ra&=&{\ts {g^2\over 2L^d}}\sum_p\la G\ra(p)
 +\sum_{r=2}^{\ell}  
  {\ts \left({ig\over \sqrt{L^d}}\right)^r}\!\!\!\!
  \sum_{p_2\cdots p_r\ne k\atop p_i\ne p_j} 
  \la G\ra(p_2)\cdots \la G\ra (p_r)\,
  \la\gamma_{k-p_2}\gamma_{p_2-p_3}\cdots 
  \gamma_{p_r-k}\ra_2  \nonumber\\
 & & \\
 \la\pi_q\ra&=&-{\ts {1\over 2}}\sum_{r=2}^\ell 
  {\ts \left({ig\over \sqrt{L^d}}\right)^r} \sum_{s=3}^{r-1} 
  \sum_{p_2\cdots p_r\ne p_2+q\atop p_i\ne p_j} \!\!\!
 \Bigl( \delta_{q,p_{s+1}-p_s}\, \la G\ra(p_2)\cdots \la G\ra(p_r)\,
   \la G\ra(p_2+q)\times  \nonumber \\
 & &\phantom{mmmmmmmmmmmmmmmmm  }
  \la \gamma_{p_2-p_3}\cdots 
   \widehat{\gamma}_{p_s-p_{s+1}}  \cdots \gamma_{p_{r-1}-p_r} 
   \gamma_{p_r-p_2-q}\ra_2 \Bigr)  \nonumber\\ 
 & &  
\eeqn
In the last line we used that $\gamma_q$ in (4.64) cannot 
 contract to $\gamma_{p_2-p_3}$ or to $\gamma_{p_r-p_2-q}$. 
 If the expectations of the $\gamma$-fields on the right hand side of 
 (4.71,72) are computed according to (4.66), one obtains the expansion 
 into diagrams. The graphs contributing to $\la \sigma_k\ra$ 
have exactly one string of 
 particle lines, each line having $\la G\ra$ as propagator, 
  and no particle loops (up to the tadpole diagram). 
 Each squiggle corresponds to a factor $\la |\gamma|^2\ra$. 
 The diagrams contributing to $\la \pi\ra$ have exactly one 
 particle loop, the propagators being again the interacting 
 two point functions, $\la G\ra$ for the particle lines and 
 $\la|\gamma|^2\ra$ for the squiggles. 
 In both cases there are no two-legged subdiagrams. 
 However, although the equation 
 $\la |\gamma_q|^2\ra={1\over 1+\la\pi_q\ra}$ resums ladder 
 or bubble  diagrams 
 (which is apparent from (4.57) or (4.26)) and 
 more general four-legged particle subdiagrams if the terms for $r\ge 4$ 
  in (4.72) are taken into account, the right hand side 
 of (4.71,72) still contains diagrams with four-legged particle 
 subdiagrams. Thus, the resummation of four-legged particle subdiagrams 
 is only partially through the complete resummation of two-legged 
 squiggle diagrams.    Also
  observe that, in going from (4.68,69) to (4.70-72), 
  we cut off the $r$-sum at some 
 fixed order $\ell$ independent of the volume since we can only expect 
 that the expansions are asymptotic ones, 
   compare the discussion in section 2. 

\goodbreak
%
%
%
%

\bigskip
\bigskip
\bigskip
\section{Concluding Remarks}

\bigskip
In the general case, without making the approximation (4.65), we
expect the following picture for a generic quartic field theoretical 
 model. 
Let $G$ and $G_0$ be the interacting and free particle Greens
function (one solid line goes in, one solid line goes out), and let 
$D$ and $D_0$ be the interacting and free interaction Greens 
function (one wavy line goes in, one wavy line goes out). 
Then we expect the following {\bf closed} set of integral equations for $G$ and $D$:
$$ G={1\over G_0^{-1}+\sigma(G,D)},\;\;\;\;\; 
  D={1\over D_0^{-1}+\pi(G,D)}   \eqno(5.1)$$
where $\sigma$ and $\pi$ are the sum of all two legged diagrams 
without two legged (particle and wavy line) subdiagrams with 
propagators $G$ and $D$ (instead of $G_0$, $D_0$). Thus (5.1) 
simply eliminates all two legged insertions by substituting them by 
the full propagators. For the Anderson model $D=D_0=1$ 
and (5.1) reduces to (2.27,35).

\par
A variant of equations (5.1) has been derived on a more 
heuristic level in [CJT] and [LW].  
Their integral equation (for example equation (40) of [LW]) reads 
$$ G={1\over G_0^{-1}+\tilde\sigma(G,D_0)}\eqno (5.2)$$
where $\tilde \sigma$ is the sum of all two legged diagrams 
without two legged particle insertions, with propagators $G$ 
and $D_0$. Thus this equation does not resum two legged 
interaction subgraphs (one wavy line goes in, one wavy line 
goes out). However resummation of these diagrams corresponds 
to a partial resummation of four legged particle subgraphs 
(for example the second equation in (5.4) below resums 
bubble diagrams), and is necessary in order to get the 
right behaviour, in particular for the many-electron system.

\par
Another popular way of eliminating two legged subdiagrams (instead 
of using integral equations) is the use of counterterms. 
The underlying combinatorial identity is the following one. 
Let 
$${\cal S}(\psi,\bar\psi)=
  \int dk \bar\psi_k G_0^{-1}(k) \psi_k+ 
  {\cal S}_{\rm int}(\psi,\bar\psi)  \eqno (5.3)   $$
be some action of a field theoretical model and let 
$T(k)=T(G_0)(k)$ be the sum of all amputated two legged 
particle diagrams without two legged particle subdiagrams, 
evaluated with the bare propagator $G_0$. Let 
$\delta{\cal S}(\psi,\bar\psi)=
  \int dk \bar\psi_k T(k) \psi_k$. Consider the model with 
action ${\cal S}-\delta{\cal S}$. Then a $p$-point function 
of that model is given by the sum of all $p$-legged 
diagrams which do not contain any two legged particle 
subdiagrams, evaluated with the bare propagator $G_0$. 
In particular, by construction, the two point function of that
model is exactly given by $G_0$. Now, since the quadratic 
part of the model under consideration 
(given by the action ${\cal S}-\delta{\cal S}$) should be 
given by the  bare Greens function $G_0^{-1}$ and the 
interacting Greens function is $G$, one is led to the 
equation $G^{-1}-T(G)=G_0^{-1}$ which coincides 
with (5.2).  
\par
Since the quantities $\sigma$ and $\pi$ in (5.1) are not 
explicitely given but merely are given by a sum of diagrams, 
we have to make an approximation in order to get a 
concrete set of integral equations which we can deal with. 
That is, we substitute $\sigma$ and $\pi$ by its lowest order 
contributions which leads to the system      
$$ G(k)={1\over G_0(k)^{-1}+\int dp\, D(p) G(k-p)},\;\;\;
  D(q)={1\over D_0(q)^{-1}+\int dp\, G(p) G(p+q)}  
\eqno(5.4)$$
This corresponds to the use of (1.3,4) retaining 
only the $r=2$ term. Thus we assume that  the expansions 
for $\sigma$ and $\pi$ are asymptotic. A rigorous proof of that 
is of course a very difficult mathematical problem and this has 
not been adressed in this paper. Roughly one may expect this if
each diagram contributing to $\sigma$ and $\pi$ allows 
a $const^n$ bound (no $n!$ and of course no divergent 
contributions). One may look in [FKLT] for an 
outline of proof for the many electron system with an anisotropic 
dispersion relation. In that case actually one obtains a 
series with a small positive radius of convergence instead of 
only an asymptotic one (because the model is fermionic), which 
simplifies the proof considerably. 

\par 
Finally we remark that  the equations (5.4) can be 
found in the literature. Usually 
they are derived from the Schwinger-Dyson equations which 
is the following {\bf non closed} set of two equations for the three 
unknown functions $G,D$ and $\Gamma$, $\Gamma$ being the
vertex function (see, for example, [AGD]):
\setcounter{equation}{4}
\beqn
G(k)&=&G_0(k)+G_0(k)
  \int dp\, G(p) D(k-p) \Gamma(p,k-p)\; G(k) \nonumber\\
D(q)&=& D_0(q)+D_0(q)\int dp\, G(p) G(p+q)
   \Gamma(p+q,-q)\; D(q) 
\eeqn
The function $\Gamma(p,q)$ corresponds to 
an off-diagonal inverse 
matrix element as it shows up for example in (4.22). Then 
application of (1.4) transforms (5.5) into (5.1). One may say 
that although the equations (5.4) are known, usually they are 
not really taken seriously. For our opinion this is due to 
two reasons. First of all these equations, being highly nonlinear, 
are not easy to solve. In particular, for models involving 
condensation phenomena like  superconductivity or 
Bose-Einstein condensation, it seems to be apropriate to 
write them down in finite volume since some quantities 
may become
macroscopic. And second, since they are usually derived 
from (5.5) by putting $\Gamma$ equal to 1 (or actually -1, 
by the choice of signs in (5.5)), one may feel pretty 
suspicious about the validity of that approximation. The 
equations (5.1) tell us that this is a good approximation if 
the expansions for $\sigma$ and $\pi$ are asymptotic. 

\medskip 
The applications of the method shown in this paper basically 
confirmed the common expectations for the particular models, 
thus one may say there are no really new results. However, we think 
it is fair to say that the computation of field theoretical 
correlation functions is an extremely difficult mathematical 
problem and therefore one should have welcome 
everything which sheds 
some new light on these problems. We hope that we could 
convince the reader that the method presented in this paper 
definitely does this.

\goodbreak

\bigskip
\bigskip
\bigskip
\noindent{\Mittel References}   \nobreak
\bigskip \nobreak
\bitem
\item[{[AG]}]  M. Aizenman, G.M. Graf, {\it Localization 
  Bounds for an Electron Gas}, Journ. Phys. A, Bd. 31, S. 6783, 1998.
\item[{[AGD]}] A. A. Abrikosov, L. P. Gorkov, I. E. 
  Dzyaloshinski, {\it Methods of Quantum Field Theory 
  in Statistical Physics}, Prentice-Hall, 1963.
\item[{[B]}] N. N. Bogoliubov, {\it Lectures on Quantum Statistics}, Vol.2, 
    Gordon and Breach, 1970
\item[{[BFS]}] V. Bach, J. Fr\"ohlich, and I. M. Sigal, 
  {\it Renormalization group analysis of spectral problems 
      in quantum field theory}, 
  Adv. in Math. 137, 205-298, 1998.
\item[{[CFS]}]  T. Chen,  J. Fr\"ohlich,  M. Seifert, {\it Renormalization Group 
  Methods: Landau-Fermi Liquid and BCS Superconductor}, Proceedings of the 
 Les Houches session {\it Fluctuating Geometries in Statistical Mechanics 
 and Field Theory}, eds. F. David, P. Ginsparg, J. Zinn-Justin, 1994. 
\item[{[CJT]}] J. M. Cornwall, R. Jackiw, E. Tromboulis, 
  {\it Effective Action for Composite Operators}, Phys. Rev. D, 
  vol.10, no.8, p.2428-2445, 1974.
\item[{[F]}] J. Fr\"ohlich, {\it On the Triviality of $\lambda\vp_d^4$ Theories 
  and the Approach to the Critical Point}, Nucl. Physics B200, 281-296, 1982. 
\item[{[FFS]}] R. Fernandez, J. Fr\"ohlich, A. Sokal, {\it Random Walks, 
    Critical Phenomena and Triviality in Quantum Field Theory}, Texts and Mongraphs 
    in Physics, Springer 1992.
\item[{[FKLT]}] J. Feldman, H. Kn\"orrer, D. Lehmann, E.
  Trubowitz,  {\it Fermi Liquids in Two Space Dimensions}, 
  in {\it Constructive Physics}, 
   Springer Lecture Notes in Physics, Bd. 446, 1994.
\item[{[FKT]}]  J. Feldman, H. Kn\"orrer, E. Trubowitz, {\it Mathematical Methods of 
  Many Body Quantum Field Theory}, Lecture Notes, ETH Z\"urich.
\item[{[FMRT]}] J. Feldman, J. Magnen, V. Rivasseau, E. Trubowitz, {\it Ward Identities 
 and a Perturbative Analysis of a U(1) Goldstone Boson in a Many Fermion 
 System}, Helvetia Physica Acta 66, 1993, 498-550.
\item[{[FT]}] J. Feldman, E. Trubowitz,  {\it Perturbation Theory for Many Fermion 
   Systems}, Helvetia Physica Acta 63,  p.156-260, 1990; 
     {\it The Flow of an Electron-Phonon 
    System to the Superconducting State}, Helv. Phys. Acta 64, p.214-357, 1991.
\item[{[K]}] A. Klein, {\it The Supersymmetric Replica Trick and Smoothness 
  of the Density of States for Random Schr\"odinger Operators}, Proceedings of 
 Symposia in Pure Mathematics, vol. 51, part 1, 1990.
\item[{[L1]}]  D. Lehmann, {\it The Many-Electron System in the Forward, 
 Exchange and BCS Approximation}, Comm. Math.  
Phys. 198, 427-468, 1998.
\item[{[L2]}] D. Lehmann, {\it The Global Minimum of the Effective Potential 
  of the Many-Electron System with
   Delta-Interaction}, to appear in Rev. 
  Math. Phys. Vol. 12, No 9, Sept. 2000.
\item[{[LW]}] J. M. Luttinger, J. C. Ward, {\it Ground State 
  Energy of a Many-Fermion System II}, Phys. Rev. vol. 118, 
  no. 5, p.1417-1427, 1960.
\item[{[MPR]}] J. Magnen, G. Poirot, V. Rivasseau, 
  {\it Ward Type Identities for the 2D Anderson Model at Weak Disorder}, cond-mat/9801217; 
  {\it The Anderson Model as a Matrix Model}, Nucl. Phys. B (Proc. Suppl.) 58, S.149, 1997. 
\item[{[P]}] G. Poirot, {\it Mean Green's Function of the 
  Anderson Model at Weak Disorder with an Infrared Cutoff}, cond-mat/9702111
\item[{[R]}] V. Rivasseau, {\it From Perturbative to Constructive
   Renormalization}, Princeton Univ. Press 1991.
\item[{[W]}]  Wei-Min Wang, {\it Supersymmetry, Witten Complex and Asymptotics for 
  Directional Lyapunov Exponents in $\mathbb Z^d$}, mp-arc/99-355; {\it Localization 
  and Universality of Poisson Statistics Near Anderson Transition}, mp-arc/99-473.

\eitem

\end{document}